\documentclass{article}
\usepackage{arxiv}

\usepackage{amsmath,amssymb,amsthm,amsfonts}

\usepackage{graphics}

\usepackage{array,subeqnarray}
\usepackage{wrapfig}
\usepackage{lineno}
\usepackage{subfig}
\usepackage{stmaryrd}
\usepackage{siunitx}
\usepackage{multirow}

\usepackage{boldfonts}
\usepackage{symbols}

\usepackage{tikz, xcolor}
\usetikzlibrary{shapes.geometric}
\usetikzlibrary{shapes,arrows}
\tikzstyle{block} = [rectangle, draw, text centered, rounded corners,
                     text width=9em,
                     minimum height=3em, node distance=1.cm]
\tikzstyle{block1} = [rectangle, draw, text centered, 
                     minimum width=2em,
                     minimum height=2em, node distance=1.cm]
\tikzstyle{line} = [draw, -latex']
\usetikzlibrary{positioning}
\ifpdf
  \DeclareGraphicsExtensions{.eps,.pdf,.png,.jpg}
\else
  \DeclareGraphicsExtensions{.eps}
\fi

\usepackage[utf8]{inputenc}
\usepackage[english]{babel}
\newtheorem{theorem}{Theorem}[section]
\newtheorem{remark}[theorem]{Remark}

\newcommand{\bfeps}{\boldsymbol{\epsilon}}

\newcommand{\bfsig}{\boldsymbol{\sigma}}

\newcommand{\curl}{\text{curl}}   				%
\newcommand{\tr}{\text{tr}}       				%

\DeclareMathAlphabet{\mathpzc}{OT1}{pzc}{m}{it}

\newcommand{\bfw}{\boldsymbol{w}}	

\title{Nonlinear Strain-limiting Elasticity for Fracture Propagation with Phase-Field Approach}

\author{ Sanghyun Lee \\ 
Department of Mathematics \\ Florida State University \\ Tallahassee, FL \\
  \texttt{lee@math.fsu.edu} 
 \And
Hyun Chul Yoon \\
Department of Mathematics \& Statistics \\
Texas A\&M University-Corpus Christi \\
Corpus Christi, TX \\
 \texttt{hyun.yoon@tamucc.edu}
\And 
Mallikarjunaiah S. Muddamallappa\\
Department of Mathematics \& Statistics \\
Texas A\&M University-Corpus Christi \\
Corpus Christi, TX \\
 \texttt{M.Muddamallappa@tamucc.edu}
}

\begin{document}

\maketitle

\begin{abstract}
The conventional model governing the spread of fractures in elastic material is formulated by coupling linear elasticity  with deformation systems. 
The classical {linear elastic fracture mechanics (LEFM)} model is derived based on the assumption of {small} strain values. 
However, since the strain values {in the model} are linearly proportional to the stress values, the strain value can 
{be large} if the stress value {increases.} 
Thus this results 
in the contradiction of the assumption to {LEFM} 
and it is one of the major disadvantages {of the model}.
In particular, this {singular} behavior of the strain values is often observed especially near the crack-tip, 
and it may not accurately predict realistic phenomena. 
Thus, we 
{investigate} the framework of a new class of theoretical model, which is known as the nonlinear strain-limiting model.
The advantage of the nonlinear strain-limiting models over LEFM is that the strain value remains bounded even if the stress value tends to the infinity. 
This is achieved by assuming the nonlinear relation between the strain and stress in the derivation of the model.
Moreover, we consider the quasi-static fracture propagation 
{by} coupling with 
the phase-field approach to present the effectiveness of the proposed strain-limiting model. 
Several numerical examples to evaluate and validate the performance of the new model and algorithms are presented. 
Detailed {comparisons} of the strain values,  fracture energy,  and fracture propagation speed between nonlinear strain-limiting model 
{and} LEFM for the quasi-static fracture propagation are discussed. 
\end{abstract}


\section{Introduction}

Fracture mechanics has been one of the major interests of the research in several different areas such as 
civil engineering, mechanical engineering, environmental engineering, petroleum engineering, and applied mathematics. Initially, Griffith \cite{griffith1921phenomena} gave some solid foundation for the energy-based brittle fracture theory by relating the energy balance between {the} stored elastic energy of the material with that of the energy required to create {a new} crack increment. 
The mathematical problems of material failure or fracture were conventionally modeled within the framework of linear elastic fracture mechanics (LEFM) which has been one of the most successful theories of applied mechanics. 
LEFM is derived based on the assumption of uniform infinitesimal strains and there by reduces to a linear relationship between Cauchy stress and strain tensors.

However, the linear relation in LEFM contains a noticeable inconsistency such that the strain values can 
proportionally rise if the stress value increases, and this contradicts the assumption of the model: the small strain.
In particular, this behavior often arises in the vicinity of crack-tip, where LEFM predicts singularity in stress and strain values~\cite{broberg1999cracks}.
Many studies have been attempted to correct these inconsistencies by augmenting LEFM based on different modeling paradigms, 
such as cohesive zone, process zone models \cite{broberg1999cracks,Barenblatt1962}, or surface mechanics based theories 
\cite{zemlyanova2013, zemlyanova2012, walton2014, sendova2010, ferguson2015, WaltonMalli2016}. 
Other modeling procedures include avoiding the crack-tip singularity in the neighborhood of the crack-tip by introducing 
surface elasticity, inelasticity, or plasticity \cite{broberg1999cracks, kanninen1985, bell1985, gurtin1975, kim2010, antipov2011}. However, all the aforementioned methods are based on introducing a separate conceptual zone 
or utilizing different models.

{Furthermore, 
there are {some} clear experimental {evidence} that {certain materials such as} titanium alloy manifests the nonlinear behavior well within the small strain regime \cite{zhang2009fatigue, saito2003multifunctional, hou2010nonlinear, withey2008deformation}. There is a recent surge in the material-science community to develop new titanium alloys due to their 
high yield strength and strain with 
{small} values of Young's modulus. These new metals have very different properties than the conventional ones, however classical linear models cannot describe the nonlinear stress response even when strains are only around $2\%$ \cite{devendiran2017thermodynamically, kulvait2017modeling}. 
Hence, it is important to provide and study some  new class of elasticity models that can capture the stress-strain response of such nonlinear materials.}

Recently, a new class of nonlinear theoretical model, which is derived based on the 
 implicit relationship between Cauchy stress and Cauchy-Green tensor has been introduced in \cite{rajagopal2003implicit, rajagopal2007elasticity, rajagopal2007response}.
Based on the implicit relationship and appealing to the standard linearization process under the assumption that the norm of  
the displacement gradient is small, one can arrive at a non-customary nonlinear relationship between the linearized strain and Cauchy stress tensor. 
Structured on this nonlinear relation, 
the strain value remains bounded even if the stress value tends to the infinity.
Such a class of nonlinear models is known as the nonlinear strain-limiting model \cite{rajagopal2011non,rajagopal2011conspectus,rajagopal2014nonlinear}. 
Rigorous mathematical analyses to show the existence of weak solutions for the variety of problems formulated within the implicit theory of elasticity are shown in \cite{bulivcek2014elastic, Bulicek2015, bulivcek2015analysis}. Convergence and analysis of the numerical schemes for crack problems are described in  \cite{bonito2018finite,gelmetti2018spectral}.
Moreover, the responses of elastic bodies \cite{bustamante2009some, bustamante2010note, bustamante2011solutions}, electro-elastic bodies \cite{bustamante2013new}, magneto-elastic bodies \cite{bustamante2015implicit}, thermo-elastic bodies \cite{bustamante2017implicit}, 
by employing  the nonlinear elasticity within general strain-limiting theory are presented in previous studies. 

In this study, we focus on coupling 
the strain-limiting model with a phase-field approach 
to investigate fracture propagation.
Recently, the phase-field approach has become a powerful tool for modeling 
fracture propagation.
In particular, the phase-field formulation derived from variational theory have received a lot of attention from the applied mechanics community due to its strong ties to Griffith's theory for brittle fracture \cite{Gri21,Barenblatt196255}.
 The advantages of this approach include the ability for automatically determining  
the direction of crack propagation, joining, and branching through minimization of an energy functional without  additional constitutive rules or criteria.
Thus,  computing  stress intensity factors near the crack-tip is intrinsically embedded in the model. 
In addition, all computations are performed entirely on the initial, un-deformed configuration. 
Therefore, there is no need to disconnect, eliminate, move elements or introduce additional discontinuity. 
This result in a significant simplification of the numerical implementation
to handle realistic heterogeneous properties of {solid or} porous media with adaptive mesh refinement in {two and three dimensional} applications. 
{Further recent} advances and numerical studies for treating multiphysics phase-field fractures include the
followings: thermal shocks and thermo-elastic-plastic solids
\cite{BouMarMauSics14,Miehe2015449,noii2019phase},
elastic gelatin for wing crack formation \cite{LeeRebHayWhe_2016}, pressurized fractures
\cite{MiWheWi15b,WiLeeWhe15},
fluid-filled (i.e., hydraulic) fractures
\cite{MiWheWi14,LeeWheWi16,MieheMauthe2015,Heider2016,LeeMikWheWick2017_pftwo,LeeWheWi17},
proppant-filled fractures \cite{LeeMiWheWi16}, variably saturated porous media \cite{Cajuhi2017},
crack initiations with microseismic probability maps \cite{LeeWheWiSri16,wheeler2019unconventional}, and many other applications \cite{choo2018cracking,Bourdin2018,Yoshioka2019,Ambati2015,SHOVKUN201942,LeeMinWhe2018,mandal2019phase,SogoLeeWheeler_2018}.   

{Therefore, we utilize the phase-field approach for the fracture propagation and couple with the nonlinear strain-limiting model. 
{Also in} this paper, we employ an iterative coupling algorithm, so-called staggered L-scheme \cite{brun2019iterative}.} Recently developed L-scheme provides an efficient iterative coupling between the phase-field and elasticity. 
An adaptive mesh refinement to localize the mesh refinement near the thin fractures is applied for the efficiency of the algorithm as in \cite{HeWheWi15}. 
Several numerical simulations are illustrated to compare the convergence of the iterative solvers, stress/strain values, and the fracture propagation, between {LEFM} and new nonlinear strain-limiting elasticity.
{In summary}, 
the main novelty of this study is to extend the strain-limiting theory to consider quasi-static fracture initiation and propagation. 
Thus, a new computational framework of formulating a quasi-static strain-limiting fracture  by iteratively coupling the nonlinear strain-limiting model with the phase-field approach is established. 

The organization of the paper is as follows: In Section~\ref{sec:background}, we briefly introduce the derivation of strain-limiting model and recapitulate the main idea of phase-field approach. 
Moreover, the mathematical models and governing system for our problem is discussed. 
Spatial and temporal discretization using finite element method and the solution algorithm is presented in Section~\ref{sec:num}.
Finally, several numerical examples comparing the classical linear elasticity model and nonlinear strain-limiting model including the quasi-static fracture propagation are illustrated in Section~\ref{sec:examples}.

\section{Mathematical Model} 
\label{sec:background}

In this section, a brief overview of the physical modeling including  
the nonlinear strain-limiting elasticity and the phase-field approach is presented 
with rationale based on previous studies. 
We first introduce the kinematical setting and notations that we use for LEFM and the nonlinear strain-limiting model.

\subsection{Strain-limiting theories for elasticity}

The strain-limiting model for elasticity which was established and discussed in \cite{rajagopal2003implicit,rajagopal2007elasticity,rajagopal2011conspectus,rajagopal2007response,rajagopal2009class} is briefly described
in this section. 
Let $\bx:= f(\bX, t)$ denote the current position of a particle (motion of a particle) that is at $\bX$ of a material body $\mathcal{A}$ in 
{the} stress-free reference configuration.

Here $f$ is a deformation of the body which is differentiable and the displacement is denoted by 
$\bu := \bx - \bX$.
Then the displacement gradients are defined as
\begin{equation}
\dfrac{\partial \bu}{\partial \bX} := \nabla_\bX \bu = \bF - \bI 
\ \ 
\text{ and } 
\ \
\dfrac{\partial \bu}{\partial \bx} := \nabla_\bx \bu =  \bI - \bF^{-1}, 
\end{equation}
where $\bI$ is the identity matrix and $\bF$ is the deformation gradient 
\begin{equation}
\bF := \dfrac{\partial f}{\partial \bX}.
\end{equation}
The left and right Cauchy-Green stretch tensors $\bB$ and $\bC$ are given by 
\begin{equation}
\text{(left) } \ \bB := \bF \bF^{\text{T}}, \ \  \text{(right) } \  \bC :=  \bF^{\text{T}} \bF, 
\end{equation}
respectively. 
Then the Green-St.Venant strain tensor $\bE$ and the Almansi-Hamel strain $\be$ are defined as
\begin{equation}
\bE := \dfrac{1}{2} (\bC - \bI) \ \ \text{ and } \ \
\be := \dfrac{1}{2} (\bI - \bB^{-1}). 
\end{equation}

\subsubsection{The linearized theory of elasticity for isotropic bodies}

Let $\bsigma$ denote the Cauchy stress tensor in a deformed configuration, then the first and second Piola-Kirchhoff {stress tensors} in a reference configuration are  
\begin{equation}
\bS:= \bsigma \bF^{-\text{T}} \text{det}(\bF) 
\ \text{ and }\ 
\bar{\bS}:= \bF^{-1} \bS,
\end{equation}
respectively. 
The material body $\mathcal{A}$ is called Cauchy elastic if its constitutive class is determined by a scalar function of the deformation gradient, i.e., 
\begin{equation} 
\bS = \hat{\bS}(\bF).
\end{equation}
Thus, the Cauchy stress $\bsigma$ is a function of the deformation gradient $\bF$, 
and the stress depends on the stress-free and final configurations of the body \cite{Truesdell2004}. 
For a compressible homogeneous isotropic Cauchy elastic body, the constitutive relation~\cite{Truesdell2004} is 
\begin{equation}
\bsigma = \alpha_1 \bI + \alpha_2 \bB + \alpha_3 \bB^2,
\label{eqn:12}
\end{equation}
where $\alpha_i$, $i=1,2,3$ depend on isotropic invariants of $\rho, \tr (\bB), \tr (\bB^2),$ and $\tr (\bB^3)$, where $\rho$ is the density of the body, and $\tr(\cdot)$ is the trace operator.
Next, the body $\mathcal{A}$ is called Green elastic (or hyper-elastic)  \cite{truesdell1955hypo}  
if the stress response function is the gradient of a scalar valued potential, i.e.,
\begin{equation} 
\hat{\bS}(\bF) = \partial_\bF \hat{w}(\bF),
\end{equation}
and hence a stored energy, $\hat{w}(\bF)$, exists.
Thus, the stress in a Cauchy elastic body and the stored energy associated with a Green elastic body depend only on the deformation gradient
as discussed in \cite{carroll2009must}. 

\subsubsection{Implicit and strain-limiting constitutive models}
\label{sec:nonlinear}

However, {the} more general class of elastic materials than Cauchy elastic bodies, 
which  assumes that the stress and the deformation gradient are related by implicit constitutive relations, is introduced by Rajagopal in \cite{rajagopal2007elasticity,rajagopal2003implicit}.
A special subclass of these implicit models is where an explicit representation is given for the left Cauchy-Green stretch tensor $\bB$ in terms of Cauchy stress $\bsigma$. These models for elastic bodies are neither Cauchy elastic nor Green elastic.

First, let us consider an isotropic implicit constitutive relation of the form 
\begin{equation}
\mathcal{F}(\bsigma, \bB) = \mathbf{0},
\label{eqn:19}
\end{equation}
between the Cauchy stress and {the} left Cauchy-Green tensor. 
Following \cite{spencer2017part}, with the assumption that the elastic body is isotropic homogeneous compressible, we obtain 
\begin{equation}
\bB = \tilde{\alpha}_1 \bI +  \tilde{\alpha}_2 \bsigma +  \tilde{\alpha}_3 \bsigma^2,
\label{eqn:23}
\end{equation}
where  $\tilde{\alpha}_i$, $i=1,2,3$ are the scalar-valued functions of the isotropic invariants of
$\rho$, $\tr(\bsigma)$, $\tr(\bsigma^2)$, and $\tr(\bsigma^3)$.
Note that the stress and the Cauchy-Green stretch are reversed compared to the classical model in Equation~\eqref{eqn:12}. 
Equation~\eqref{eqn:23} cannot be obtained from the class of general Cauchy elastic bodies by inverting the stress as a function of the deformation gradient \cite{rajagopal2007elasticity}.
Under the assumption of small displacement gradients such that, 
\textcolor{black}{\begin{equation}
\max \| \nabla_\bx \bu \| = \mathbf{0}(\delta), \ \delta \ll 1,
\label{eqn:max}
\end{equation}}
we obtain 
\textcolor{black}{\begin{equation}
\bE = \bfeps + \mathbf{0}(\delta^2), \  \ 
\be = \bfeps + \mathbf{0}(\delta^2), \  \
\bB = \bI + 2\bfeps + \mathbf{0}(\delta^2),
\label{eqn:max-2}
\end{equation}}
where $\bfeps$ is the linearized 
strain: 
\begin{equation}\label{LinStrain}
\bfeps  
:= \bfeps(\bu) 
= \dfrac{1}{2}\left(  \nabla \bu + (\nabla \bu)^\text{T} \right).
\end{equation}

Finally, the  linearization of the model, Equation~\eqref{eqn:23}, under the assumption of small displacement gradient~(Equation~\eqref{eqn:max}-\eqref{eqn:max-2}) leads to 
\begin{equation}
\bfeps = \beta_1 \bI + \beta_2 \bsigma + \beta_3 \bsigma^2, 
\label{eqn:30}
\end{equation}
where the linearized strain is given as a nonlinear function of Cauchy stress and here the $\beta_1$ is dimensionless coefficient and
 material moduli $\beta_2$ and $\beta_3$ need to have dimensions that 
are the inverse of the stress and the square of the stress, respectively.

The above approximation, Equation~\eqref{eqn:30}, has no restrictions on the stress while requiring that the strain to be small. 
This nonlinear relationship could be crucial, since one could have bounded (limiting) strains 
even if the non-dimensional stress tends to a large value.
Such models 
have very interesting applications, particularly dealing with crack and notch problems, {which} within  classical linearized elasticity may lead to 
unrealistic singular strains, but the model (Equation~\eqref{eqn:30}) predicts physically reasonable strains.
Clearly, applications including cracks and fracture in elastic bodies 
are one of the {areas} but it is not limited to those problems.

\begin{remark}
Under the assumption of Equation~\eqref{eqn:max}, we note that there is no distinction between $\bE,  \be$ and $\bfeps$, and we do not distinguish between reference and deformed configurations for linear elastic materials.
\end{remark}

\begin{remark}
For the isotropic linear elastic material in the absence of body force, 
the linear and angular momentum balance reduces to 
\begin{equation}\label{eq:blaws}
-\nabla \cdot \bsigma = \bf0, \quad \bsigma = \bsigma^T.
\end{equation}
If the displacement (including in the neighborhood of stress concentrators as crack-tips, {reentrant} notch-tips, etc.) is smooth enough, one can consider formulating boundary value problem using \eqref{eq:blaws}. Further, the linearized strain tensor needs to satisfy the compatibility conditions such as
\begin{equation}\label{eq:compcond}
\curl \, \curl \, \bfeps = \bf0, 
\end{equation}
where $\curl$ is the classical operator for tensors. In the view of Equation~\eqref{eq:blaws}, Equation~\eqref{eq:compcond} will be automatically satisfied for {a} linear elastic material. 
\end{remark}

For isotropic, homogeneous, linear elastic material, the constitutive relationship {for Cauchy stress} is given by {Hooke's law:}
\begin{equation}\label{eq:HLaw}
\bfsig = 2\mu \, \bfeps  + \lambda \, \tr \left( \bfeps \right) \, \bI,
\end{equation}
where 
$\mu$ and $\lambda$ are Lam$\acute{e}$ parameters and $\tr(\cdot)$ is the trace operator for tensors. 
Since  Equation~\eqref{eq:HLaw} is invertible, we can express linearized strain tensor {\color{black}$\bfeps$}
as a (linear) function of Cauchy stress as
\begin{equation}\label{eq:HLawInv}
\bfeps= \dfrac{1}{2\mu} \, \bfsig  - \dfrac{\lambda}{ 6 \mu ( \lambda+ (2/3) \mu)}  \, \tr \left( \bfsig \right) \, \bI.
\end{equation}
Hence, one can formulate the boundary value problems for linear elastic material either within Equation~\eqref{eq:HLaw} or Equation~\eqref{eq:HLawInv}. However, as a result,  the strains in the neighborhood of crack-tips will be large, which clearly violates the fundamental assumption,
(as assumed in Equation~\eqref{eqn:max}), as a consequence of which the theory of linear elastic materials was derived. In general, for most elastic materials, the shear modulus $\mu$ is positive and the term $\lambda + (2/3) \,  \mu$, in Equation~\eqref{eq:HLawInv},  called the bulk modulus (which has the same units as stress), 
can never be zero.

Now, let's consider the special subclass of strain-limiting constitutive relationship from Equation~\eqref{eqn:30}, having the form as
\begin{equation}
\bfeps = \Psi_{0}\left(\text{tr}(\bsigma),  |\bsigma|  \right) \bI + \Psi_{1}\left( |\bsigma|  \right) \bsigma,
\label{eqn:main}
\end{equation}
and which is generally non-invertible. In the above Equation~\eqref{eqn:main}, $\Psi_{0}(\cdot, \cdot),  \Psi_{1}(\cdot)$ 
are scalar functions of stress invariants and more importantly the assumption of no residual stress implies $\Psi_{0}\left( 0,  \cdot \right) = 0$. 

In this paper, we extend  these previous frameworks for static cracks to quasi-static crack evolution by considering special subclass of nonlinear models that are invertible, yet rank-one convex and does not loose strong ellipticity. To that end, let us consider a nonlinear, hyperelastic model in the infinitesimal strain regime 
as:
\begin{equation}
\bE = \phi (\mathbb{K}[\bar{\bS}]) \mathbb{K}[\bar{\bS}],
\label{eqn:E_aniso}
\end{equation}
where $\mathbb{K}$ is the compliance tensor, $\bE$ is the Green-Lagrangian strain, and {$\bar{\bS}$ is the second Piola-Kirchhoff stress}.
Note that the above model can be customized for an anisotropic material model and it was shown in \cite{MaiWalton} that the models of the type, Equation~\eqref{eqn:E_aniso}, fail to be rank-one convex (equivalently loose the notion of strong ellipticity) if the strains are large. A simpler model to consider within the general class of models described by Equation~\eqref{eqn:E_aniso} is
\begin{equation}
\phi(\mathbb{K}[\bar{\bS}]) := \tilde{\phi}(|\mathbb{K}^{1/2}[\bar{\bS}] |),
\label{eqn:phi_1}
\end{equation}
as described in \cite{Mallikarjunaiah2015}. In Equation~\eqref{eqn:phi_1}, 
 $\tilde{\phi}(r)$ is a positive, monotonic decreasing function and 
$r\tilde{\phi}(r)$ is uniformly bounded for $0 < r< \infty$, and $\mathbb{K}^{1/2}[\cdot]$ denotes the unique, positive definite square-root of the compliance tensor $\mathbb{K}$. 
One special case of Equation~\eqref{eqn:phi_1} is defined with 
\begin{equation}\label{eqn:phi_tilde}
\tilde{\phi}(r) := \dfrac{1}{(1 + (\beta r)^{\alpha} )^{1/\alpha}},
\end{equation}
where $\alpha$ and $\beta$ are the nonlinear {model} parameters 
\cite{bulivcek2014elastic,gou2015modeling,Mallikarjunaiah2015,rajagopal2011modeling}. 
Some detailed studies of these parameters are presented in the numerical example section.
\begin{remark}
The function $\tilde{\phi}(r)$ in \eqref{eqn:phi_tilde} needs to be a decreasing function with $\beta >0$ and $\alpha>0$ for the strains to be ``limited'' near the crack-tip. Using the function $\tilde{\phi}(r)$, one can fix an upper bound for strains a priori to model specific materials or physical experiments with real data. The assumption of $\beta$ being positive is very important for the model to be hyperelastic and invertible, and the same has been observed in several other studies involving strain-limiting models \cite{bulivcek2014elastic,bulivcek2015analysis,bonito2018finite,fu2019generalized,fu2019constraint,Bulicek2015}.
\end{remark}

Thus, under the infinitesimal 
 strain {assumption}, we arrive at {\color{black} the {\textit{nonlinear relation}} between strain $\bfeps$ and stress, such as}
\begin{equation}
{\color{black} \bfeps }
=  \tilde{\phi}(|\mathbb{K}^{1/2}[\bar{\bS}] |) \mathbb{K}[\bar{\bS}],
\label{eqn:phi_2}
\end{equation}
where $\bar{\bS}$ can be viewed as {Cauchy} stress, i.e $\mathbb{K}[\bar{\bS}]=\mathbb{K}[\bsigma]$.
{From} the relation in  Equation~\eqref{eq:HLawInv}, we obtain
\begin{equation}
\mathbb{K}[\bsigma]\:{:=\bfeps=}\: \frac{\bsigma}{2\mu}-\frac{\lambda\,\tr(\bsigma) \, \bI}{2\mu(2\mu+3\lambda)},
\label{eqn:phi_4}
\end{equation}
where $\bsigma$ is obtained from Equation~\eqref{eq:HLaw}.
Finally, by using Equation~\eqref{eqn:phi_tilde}-\eqref{eqn:phi_4}, we obtain the following nonlinear relation for the strain $\bfeps$ by
\begin{equation}
{\color{black}\bfeps := }~\bfeps_\text{NL} =
 \dfrac{\mathbb{K}[\bsigma]}{(1 + (\beta |\mathbb{K}^{1/2}[\bsigma] |)^{\alpha} )^{1/\alpha}},
\label{eqn:nonIinear1}
\end{equation}
where 
$$
|\mathbb{K}^{1/2}[\bsigma] |=\left(\frac{\bsigma:\bsigma}{2\mu}-\frac{\lambda\,\tr(\bsigma)^2}{2\mu(2\mu+3\lambda)}\right)^{1/2}.
$$
We note that we are denoting the strain $\bfeps$ in two different forms depending on the formulations.  
The nonlinear strain-limiting strain ($\bfeps_\text{{NL}}$) is the same as 
{$\bfeps$} in 
Equation~\eqref{eq:HLawInv}
provided $\beta=0$ or $\alpha\rightarrow\infty$. Henceforth, unless otherwise noted, we use the notation $\bfeps_\text{{NL}}$ only for the strain obtained by the nonlinear model. 

To formulate boundary value problems   
within the framework of the new class of nonlinear models, 
we start
by  setting  the displacement ($\bu$) as the primary variable as expressed in Equation~\eqref{LinStrain}.
Here, the strain compatibility condition (Equation~\eqref{eq:compcond}) is automatically satisfied. Then, we invert Equation~\eqref{eqn:main} to get the components of stress tensor and replace  these components in Equation~\eqref{eq:blaws} to obtain a quasi-linear partial differential equation. 
Recently, it was shown in \cite{Mallikarjunaiah2015,gou2015modeling,rajagopal2011modeling,MalliPhD2015} that models within the context of Equation~\eqref{eqn:main} for the problem of a static crack in a body undergoing anti-plane shear lead to solutions with the bounded  strains at the crack-tip. 

Then, since Equation~\eqref{eqn:nonIinear1} is invertible and letting $\bar{\bS} =\bsigma$ to formulate in a deformed configuration, 
\color{black}the partial differential equation in the form of Equation~\eqref{eq:blaws} for the  proposed strain-limiting model is derived as 
\begin{equation}
-\nabla \cdot  \dfrac{\mathbb{E}[\bfeps]} 
{(1 - (\beta |\mathbb{E}^{1/2}[\bfeps]|)^\alpha )^{1/\alpha}} = \bf0.
\label{eqn:linear_pf_new}
\end{equation}
Here $\mathbb{E}$ is the fourth order linearized elasticity tensor and is symmetric and positive definite. For the isotropic, homogeneous materials, we have
\begin{equation}\label{eqn:stress}
\mathbb{E}[\bfeps]\:{:=\bfsig=}\:2 \, \mu \, \bfeps + \lambda \, \tr(\bfeps) \, \bI,
\end{equation}
by considering the displacement as the primal unknown variable, and the strain is given as the symmetric gradient of the displacement as in Equation~\eqref{LinStrain}.
The other method to directly compute the explicit nonlinear
stress by using Airy stress function is shown in \cite{ortiz2012, kulvait2013, kulvait2019}.

\begin{remark}\label{rmk001}
{From Equation~\eqref{eqn:linear_pf_new}, it} is required to satisfy the following condition,  
\textcolor{black}{\begin{equation}
\left( 1 - \left( \beta  | \mathbb{E}^{1/2} \left[ \bfeps \right]|\right)^{\alpha}\right)^{1/\alpha} > 0,
\label{eqn:beta_condition}
\end{equation}}
for the ellipticity for the weak formulation. This condition is similar to the results provided in \cite{bonito2018finite,fu2019generalized,fu2019constraint} for their analyses. 
We note that this condition reflects 
Lam\'{e} coefficients within the calculation of $| \mathbb{E}^{1/2} \left[ \bfeps \right]|$ and the choice of the nonlinear parameters, $\alpha$ and $\beta$. More detailed conditions particularly related to the strain-limiting effects are addressed in Section 4. 
\end{remark}

\subsection{Phase-field approach for fracture propagation with nonlinear strain-limiting elasticity}
\begin{figure}[!h]
\centering
\includegraphics[width=0.5\textwidth]
{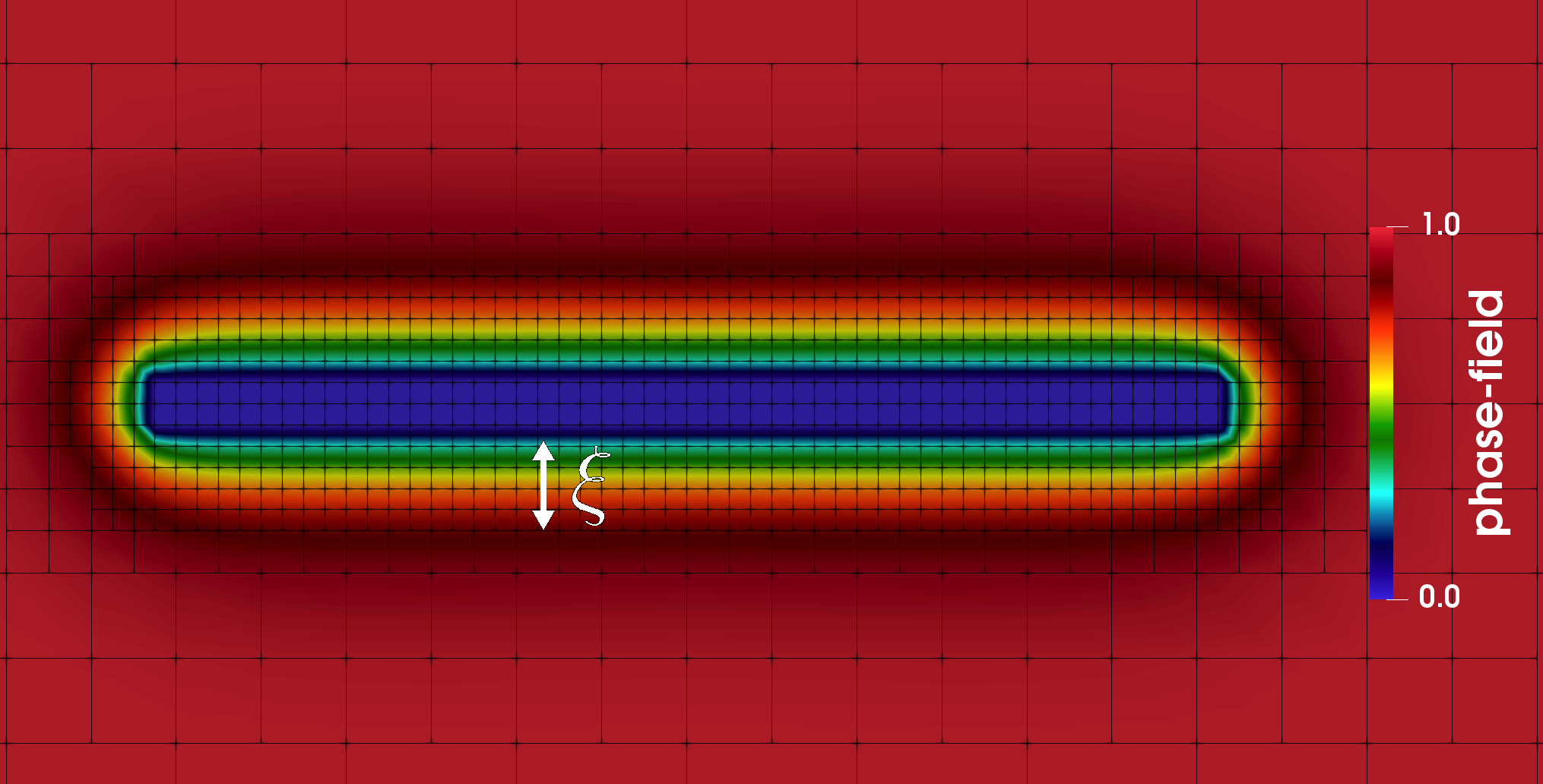}
\caption{An example of a fracture defined with the phase-field function $\varphi(\cdot, t) \in [0.1]$. }
\label{fig:figure_1}
\end{figure}

Let  $\Lambda := \Lambda(t) \in \mathbb{R}^d$ ($d=2,3$) be a smooth open and bounded computational domain, with a given boundary $\partial \Lambda$. 
Here, the time is denoted by $t \in [0, T]$,
with the final time $T>0$ in the computational time interval. 
As discussed in \cite{BourFraMar00,FraMar98}, the fracture 
$\mathcal{C}(t)$ is contained compactly in $\Lambda(t)$.
In the phase-field fracture approach, 
discontinuities in the displacement field $\bu$ across the lower-dimensional crack surface
{is} approximated by a smooth scalar function $\varphi(\cdot, t):  \Lambda \times [0,{T}] \rightarrow [0,1]$. 
This phase-field function $\varphi(\cdot, t)$ introduces
a diffusive transition zone,
which has a  bandwidth $\xi$,
between the fractured region ($\Omega_F$) having $\varphi(\cdot, t) = 0$ and the un-fractured region 
having $\varphi(\cdot, t) = 1$. See Figure \ref{fig:figure_1} for more details.
The boundary of the fracture 
is denoted by $\Gamma_F(t) := \bar{\Omega}_F(t) \cap \bar{\Omega}_R(t)$.

To discuss the phase-field fracture, 
we first introduce the Francfort-Marigo functional \cite{FraMar98}, which describes the energy with a fracture in an elastic 
{body} as  
\textcolor{black}{\begin{equation}
E(\bu, \mathcal{C}) = \dfrac{1}{2} \int_{\Omega_R}  \bsigma(\bu) : \bfeps(\bu)   \ d\bx + G_c H^{d-1}(\mathcal{C}), 
\label{eqn:energy_func}
\end{equation}}
where $\bu(\cdot,t): \Omega_R \times [0,T] \rightarrow \mathbb{R}^d$ is the solid's displacement,
$\bsigma(\bu)$  is the Cauchy stress tensor and $\bfeps(\bu)$ is the linearized strain tensor. 
Here the first term in the right-hand side is the strain energy in an un-fractured region and the second term is the fracture energy,
where the Hausdorff measure $H^{d-1}(\mathcal{C})$ denotes one dimension less fracture scale such as 
the length of the fracture in two-dimensional domain and is multiplied by 
$G_c$, i.e., the critical energy release rate. 

Next, we consider the global constitutive dissipation functional of Ambrosio-Tortorelli type \cite{AmTo90,AmTo92} to regularize the total energy with the introduction of a phase-field function.
Equation~\eqref{eqn:energy_func} is rewritten as the global dissipation formation, such as,  
\begin{equation}
E_{\xi}(\bu, \varphi) = 
\int_{\Lambda} \dfrac{1}{2} \varphi^2 \sigma(\bu) \colon \bfeps(\bu) \ d\bx 
 + G_c \int_{\Lambda} \left( \dfrac{1}{2\xi} (1-\varphi )^2 + \dfrac{\xi}{2}  | \nabla \varphi |^2 \right)\, d\bx,
\label{energy_functional}
\end{equation}
where all definitions are extended to $\Lambda$.
{Finally, we seek the solution $\bu$ and $\varphi$ which minimizes the energy functional $E_{\xi}(\bu, \varphi)$, i.e. find $\{\bu, \varphi \}$ such that 
\begin{equation}
\min_{\bu,\varphi}E_{\xi}(\bu, \varphi),
\label{energy_functional_min}
\end{equation}
{of which} approach was initially introduced for linear elasticity in \cite{BourFraMar00,FraMar98,MieWelHof10a}.}
 ~In addition, the convergence of time discrete solutions of Equation~\eqref{energy_functional_min} 
 to continuous solutions as timestep goes to zero was discussed in \cite{francfort2003existence,giacomini2005ambrosio}.
This approach becomes as a variational inequality since the fracture propagation is required to satisfy a crack irreversibility constraint, which  
is given as $\partial_t \varphi(\cdot,t) < 0$. This condition only allows the phase-field value to decrease in time and enforces the fracture to only propagate but not to heal.
The phase-field function is subject to homogeneous Neumann conditions on 
$\partial\Lambda$. 
For the quasi-static system, the initial domains, $\Omega_F(\cdot,0)$ and $\Omega_R(\cdot,0)$, are defined by a given initial phase-field value $\varphi(\cdot,0)$, either by $0$ or $1$.

{We note that the previous numerical results of phase-field approach in \cite{LeeRebHayWhe_2016,LeeWheWi16,LeeWheWi17} employ 
the classical linear elasticity {with LEFM} such as
\begin{equation}
\bsigma(\bu) = 2 \mu \bfeps(\bu) + \lambda(\nabla \cdot \bu) \bI,
\label{eqn:linear_pf}
\end{equation} 
and several numerical examples  illustrate large stress/strain values near the crack-tip. 
In this study, we extend the quasi-static fracture model to consider the nonlinear strain-limiting theory 
{addressing} the issue.}
In the associated strain energy function in Equation~\eqref{energy_functional}, 
the linear elastic stress tensor Equation~\eqref{eqn:linear_pf} will be replaced by utilizing the proposed strain-limiting model (Equation~\eqref{eqn:linear_pf_new}),
\begin{equation}
\bsigma(\bu) =  \dfrac{\mathbb{E}[\bfeps]}
{(1 - (\beta |\mathbb{E}^{1/2}[\bfeps]|)^\alpha )^{1/\alpha}},
\label{eqn:linear_pf_1}
\end{equation}
where $\alpha, \beta >0$, and  $\mathbb{E}[\bfeps]:=2 \, \mu \, \bfeps + \lambda \, \tr(\bfeps) \, \bI$.
{\color{black} In this paper, we implement both 
Equation \eqref{eqn:linear_pf} and Equation \eqref{eqn:linear_pf_1} as two different 
{models of the linear and the nonlinear} for the choice of  
 $\bsigma$ and compare the results. }

\section{Numerical Method}
\label{sec:num}

{In this section, we present the finite element method utilized for the spatial discretization with the temporal discretization to consider the quasi-static problem and the irreversibility condition.    In addition, the Euler-Lagrange formulation for our governing system and the linearization of the given nonlinear problems are discussed. Finally, the coupling between the elasticity and the phase-field equations, so-called L-scheme is presented. } 

\subsection{{Temporal discretization and augmented Lagrangian penalization}}

We define a partition of the time interval $0=:t^0 <t^1 < \cdots <  t^N := T $ and denote the uniform timestep size by $\Delta t:= t^n - t^{n-1}$. 
Then, we denote the temporal discretized solutions by 
\begin{equation}
\bu^n := \bu(\cdot, t^n) \ \ \text{ and } \ \ 
\varphi^n := \varphi(\cdot, t^n).
\end{equation}

Here, the irreversibility condition $\partial_t \varphi < 0$ is discretized by $\varphi^n \leq \varphi^{n-1}$ $(\varphi^n - \varphi^{n-1}\leq 0)$  with employing the backward Euler method. 
Due to this irreversibility condition, 
the energy minimization problem \eqref{energy_functional_min}  becomes the constrained energy minimization problem. 
Thus, now we seek for the solution $\bu^n$ and $\varphi^n$ minimizing 
\begin{equation}\label{eqn:penal}
\min_{\bu^n,\varphi^n}E_{\xi}(\bu^n, \varphi^n) + \dfrac{1}{2 \gamma}
\|[\omega_\gamma + \gamma  ( \varphi^n - \varphi^{n-1} ) ]^ + \|^2,
\end{equation} 
for each timestep $n$ with given  $\varphi^{n-1}$.
The last term is the  penalization term to enforce the irreversibility condition as discussed in \cite{wheeler2014augmented,brun2019iterative}. 
Here $\gamma > 0 $  is the penalization parameter and the choice of $\gamma$ is very sensitive to the numerical results. If $\gamma$ is too small, the irreversibility condition will not be enforced enough and if $\gamma$ is too large, the linear system becomes ill-conditioned. 
For the better performance, we utilize the augmented Lagrangian method \cite{fortin2000augmented,glowinski1989augmented,wheeler2014augmented} by  adding a function $\omega_\gamma \in L^2(\Lambda)$ which is given and updated  through the iteration. 
Moreover, here  $[\cdot]^+$ denotes the positive part of a function, i.e., $[f]^+ := \max(0,f).$

\subsection{Spatial  discretizations and Euler-Lagrange equations}
We consider continuous Galerkin finite element methods for the coupled system. A mesh family $\{\mathcal{T}_h \}_{h>0}$ is assumed to be shape regular in the sense of Ciarlet, and we 
assume that each mesh $\mathcal{T}_h$ is a subdivision of $\bar{\Lambda}$ made of disjoint elements $\mathcal{K}$, i.e., squares when  $d=2$ or cubes when $d=3$. 
Each subdivision is assumed to exactly approximate the computational domain, thus $\bar{\Lambda} = \cup_{\mathcal{K}\in\mathcal{T}_h} \mathcal{K}$. 
The diameter of an element $\mathcal{K}\in \mathcal{T}_h$ is denoted by $h$
and we denote $h_{\min}$ for the minimum. For any integer $k \geq 1$ and any
$\mathcal{K} \in \mathcal{T}_h$, we denote by $\mathbb{Q}^k(\mathcal{K})$ the
space of scalar-valued multivariate polynomials over $\mathcal{K}$ of partial
degree of at most $k$. The vector-valued counterpart of $\mathbb{Q}^k(\mathcal{K})$ is denoted 
$\pmb{\mathbb{Q}}^k(\mathcal{K})$. 
Here, we set $k=1$ to consider the piecewise linear finite elements.

Let $V_h \times W_h$ be the discrete space formulated by the continuous Galerkin approximations where 
\begin{align}
&V_h({\mathcal{T}_h}) := 
\{ W \in C^0(\bar{\Lambda};\mathbb{R}^d) \ | \ W= \boldsymbol{0} \ \text{on } \partial \Lambda, W|_{\mathcal{K}} \in \pmb{\mathbb{Q}}^1(\mathcal{K}), \forall \mathcal{K} \in {\mathcal{T}_h} \} , \\
&W_h({\mathcal{T}_h}) := 
\{ Z \in C^0(\bar{\Lambda};\mathbb{R}) | 
\ Z^{n}\leq Z^{n-1} \leq 1, Z|_{\mathcal{K}} \in \mathbb{Q}^1(\mathcal{K}), \forall \mathcal{K} \in {\mathcal{T}_h} \}.
\end{align}
The spatial 
discretized solution variables are 
$\bu_h \in \mathcal{C}^1([0,T];V_h(\mathcal{T}))$ and 
$\varphi_h \in \mathcal{C}^1([0,T];W_h(\mathcal{T}))$.
{For the simplicity of our presentation, we omit the $h$-subscript, and we only consider the discrete solutions henceforth.}

Next, we formulate the variational form of the energy functional 
$E_{\xi}(\bu^n,\varphi^n)$ in Equation~\eqref{eqn:penal}  by employing the Euler-Lagrange equations and the finite element discretizations. 
Thus, find $U^n := \{\bu^n, \varphi^n \} \in V_h \times W_h$ such that 
\begin{multline}
A(U^n)(\psi) =
(((1-{\kappa}) (\varphi^n)^2 + {\kappa})~\bsigma(\bu^n) , \bfeps(\bfw))
 - 
G_c (  \dfrac{1}{\xi}  (1-\varphi^n) , \psi) \\
 +
G_c ( {\xi} \nabla \varphi^n ,\nabla \psi) =0,  
\ \ \forall \psi \in \Psi := \{ \bfw, \psi \} \in V_h \times W_h,
\label{eqn:1}
\end{multline}
for each $t^n$. 
We note that ${\kappa}$ ($0 < {\kappa} \ll {\xi} \ll 1$) is a numerical regularization parameter depending on $h$ to ensure  the numerical stability \cite{MiWheWi15b}. For the simplicity, we define the degradation function as 
$$
g({\varphi}) := ((1-{\kappa}) (\varphi^n)^2 + {\kappa}).
$$

Then, by computing the directional derivative of Equation~\eqref{eqn:1} with respect to $\bu$ and $\varphi$, we obtain the following subproblems 
\begin{equation}
A_1(\bu^n, \bfw)
:= (g({\varphi}) ~\bsigma( \bu^n ) , \bfeps(\bfw)) = 0,
 \ \ \forall \bfw \in V_h,
\label{eqn:main1}
\end{equation}
and 
\begin{multline}
A_2(  \varphi^n, \psi)
:=
(1-{\kappa})(   \varphi^n  \bsigma (\bu^n): \bfeps(\bu) , \psi )
 -
 G_c (  \dfrac{1}{\xi} ( 1-   \varphi^n )  , \psi) \\
 +
 G_c ( {\xi} \nabla  \varphi^n ,\nabla \psi) 
 + ( [\omega_\gamma + \gamma(\varphi^n - \varphi^{n-1})]^+, \psi) =0, 
\ \ \forall  \psi \in W_h.
\label{eqn:main2}
\end{multline}   
{Here, we denote $A_1$ as the mechanics subproblem and 
$A_2$ as the phase-field subproblem.}  We note that the time-discretized system, Equation~\eqref{eqn:main1}-\eqref{eqn:main2}, was analyzed in \cite{neitzel2017optimal,mikelic2019phase} by showing the existence of one global minimizer $(\bu^n, \varphi^n) \in V_h \times W_h$.

\subsection{Newton method and {iterative} algorithm}
\label{sec:iter_algo}
In this section, we briefly recapitulate and extend the staggered L-scheme introduced in \cite{brun2019iterative} 
for iteratively coupling the mechanics subproblem (Equation~\eqref{eqn:main1}) and the phase-field subproblem (Equation~\eqref{eqn:main2}). 
For each timestep $n$, the iterative algorithm defines a sequence 
$\{ \bu^{n,i}, \varphi^{n,i} \}$, where $i=1,2,\cdots, N_i$ indicates each iteration steps.
The L-scheme iteration  for our system is formulated with two steps. 
First, the mechanics subproblem (Equation~\eqref{eqn:main1}) 
is solved with {the given phase-field and displacement values given from the previous iteration, $\{ \bu^{n,i-1}, \varphi^{n,i-1} \}$. }
For the first iteration ($i=1$), we set $\bu^{n,i-1}=\bu^{n,0}:=\bu^{n-1}$ ($\varphi^{n,i-1}=\varphi^{n,0}:=\varphi^{n-1}$).
Then, the phase-field subproblem of Equation~\eqref{eqn:main2} 
is solved with the displacement value, $ \bu^{n,i}$. 
Each nonlinear subproblem is linearized by utilizing the Newton method. 
For the faster convergence of our nonlinear problem, we note that the linear problem is employed for the initial guess for the initial iteration.

\begin{figure}[ht]
\centering
\begin{tikzpicture}[]
 \node [block1] (Init) {\footnotesize $t=t^n$};
 \node [block,right of=Init] (sol_dis) [right=0.05cm of Init]  {\footnotesize \textbf{Step 1.} Solve  \\ Displacement ($\bu$); \\ 
Equation \eqref{eqn:main1} 
 \\ (Newton Iteration)};
 \node [block,right of=sol_dis] (sol_phi) [right=0.1cm of sol_dis] {\footnotesize \textbf{Step 2.} Solve  \\ Phase-Field ($\varphi$); \\
Equation \eqref{eqn:main2}
  \\ (Newton Iteration)};
 \node [block1, right of=sol_phi] (end) [right=0.05cm of sol_phi]  {\footnotesize $t=t^{n+1}$};

\path [line] (Init) -- (sol_dis);
\path [line] (sol_dis) -- (sol_phi);
\path [line] (sol_phi) -- (end);

\draw[] (5.6,-1.3) node[above]     {\footnotesize   Augmented-Lagrangian Iteration};
\draw[] (5.6,-1.65) node[above]  {\footnotesize  \& L-scheme Iteration};

\path [line]  (2.6,-1.65) -- (2.6,-0.81);
\draw [line]  (8.6,-0.81) -- (8.6,-1.65);

\draw          (8.6,-1.65) -- (2.6,-1.65);


\end{tikzpicture}
\caption{The global iterative algorithm flowchart.}
\label{fig:arg_flow}
\end{figure}
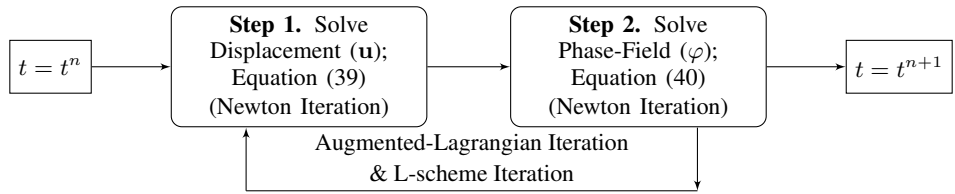

In summary, Figure \ref{fig:arg_flow} illustrates the overall global  solution algorithm for our proposed coupled system.
We note that the augmented-Lagrangian iteration to update the penalty parameter $\gamma$ and $\omega_\gamma$ is combined with the L-scheme iteration. 

\subsubsection{Step 1. Solve the mechanics subproblem for the displacement} 

In this section, we describe the details of the solution algorithm with L-scheme iteration for the mechanics subproblem to find the displacement ($\bu$). 

For each timestep $n$, and for each iteration $i$  we seek for $\bu^{n,i} \in V_h$ with given $\bu^{n,i-1},\varphi^{n,i-1}$
satisfying 
\begin{equation}
A_1(\bu^{n,i},\bfw) = 0, \ \ \forall \bfw \in V_h,
\label{eqn:MNL} 
\end{equation}
where
\begin{equation}
A_1({\bu^{n,i}}, w)
:= ( {g({\varphi^{n,i-1})}}~\bsigma( {\bu^{n,i}} ) , \bfeps(\bfw))
+
L_u(\bu^{n,i} -\bu^{n,i-1}, \bfw).
\end{equation}
Here, the last term is an additional term from the L-scheme iterative method \cite{brun2019iterative} with a given positive parameter $L_u$.

To solve  Equation~\eqref{eqn:MNL}, we employ the Newton iteration,  
and we find $\delta \bu^{n,i,a} \in V_h$ by solving
\begin{equation}
A'_1(\bu^{n,i,a-1}, {\varphi^{n,i-1}})({\delta \bu^{n,i,a}}, \bfw) = -A_1(\bu^{n,i,a-1})(\bfw), \ \ \forall \bfw \in V_h,
\end{equation}
for the Newton iteration step, $a=1,2,\cdots,$ until 
$\| \delta \bu^{n,i,a}\| \leq \varepsilon_a$.
Then the Newton update is given by 
\begin{equation}
\bu^{n,i,a} = \bu^{n,i,a-1} + \omega_u \delta \bu^{n,i,a} ,
\end{equation}
where $\omega_u$ is a line search parameter $\omega_u \in [0,1]$. 
If the Newton iteration converges, we set 
$$\bu^{n,i} = \bu^{n,i,a}.$$
Here, the Jacobian of $A_1$ is computed as
\begin{equation}\label{eqn:linear_A_prime}
A^{\prime}_1(\bu^{n,i}, {\varphi^{n,i-1,{a-1}}})({\delta \bu^{n,i,a}}, \bfw) := 
({g({\varphi^{n,i-1})}}~\bsigma( {\delta \bu^{n,i,a}} ), \bfeps(\bfw))  
+
L_u(\delta \bu^{n,i,a}, \bfw),
\end{equation}
and
\begin{equation}\label{eqn:linear_A}
A_1(\bu^{n,i,{a-1}}, \bfw)
:= ({g({\varphi^{n,i-1})}}~\bsigma( \bu^{n,i,a-1} ) , \bfeps(\bfw))  \\
+
L_u( \bu^{n,i,a-1} -\bu^{n,i-1}, \bfw).
\end{equation}

As aforementioned, here we consider two different cases for the choice of $\bsigma$. 
First, for the classical linear elasticity case, we define 
\begin{equation}\label{eqn:sig_lin}
\bsigma( {\bu^{n,i,{a-1}}}) :=  \mu \left(\nabla  {\bu^{n,i,{a-1}}} + \nabla  {\bu^{n,i,{a-1}}}^T \right) + \lambda(\nabla \cdot  {\bu^{n,i,{a-1}}}) \bI.
\end{equation}
Next, we recall the nonlinear constitutive relationship between linearized strain and Cauchy stress. 
The inverted form of stress by considering the displacement $\bu$ as the primary variable is defined as  
\begin{equation}\label{eq:nonlinear-sigma-plus}
\bsigma( \bu^{n,i,{a-1}} ):=\frac{  \mu \left( { \nabla  \bu^{n,i,{a-1}} + \left(\nabla  \bu^{n,i,{a-1}} \right)^T }  \right)  + \lambda \, (\nabla \cdot  \bu^{n,i,{a-1}})\,  \bf{I}  }{\left( 1 - \left( \beta  | \mathbb{E}^{1/2} \left[ \bfeps^{n,i,{a-1}} \right]|\right)^{\alpha}\right)^{1/\alpha}},
\end{equation}
where
\begin{align}
\left| \mathbb{E}^{1/2} \left[  \bfeps^{n,i,{a-1}}   \right] \right|^{2} &= \mathbb{E}^{1/2}[ \bfeps^{n,i,{a-1}} ] \colon \mathbb{E}^{1/2} \left[  \bfeps^{n,i,{a-1}}  \right]  \notag  \\
&= \bfeps^{n,i,{a-1}} \colon \mathbb{E}^{1/2}[\mathbb{E}^{1/2} \left[  \bfeps^{n,i,{a-1}}   \right] ] \notag  \\
&= \bfeps^{n,i,{a-1}}  \colon \mathbb{E} \left[  \bfeps^{n,i,{a-1}}   \right] \notag  \\
& = 2\mu \, \left(\dfrac{ \nabla  \bu^{n,i,{a-1}}   + \nabla  (\bu^{n,i})^{T} }{2}\right) \colon \left(\dfrac{\nabla  \bu^{n,i,{a-1}}   + \nabla  (\bu^{n,i,{a-1}})^{T}}{2}\right)   +  \lambda\, \left( \nabla \cdot   \bu^{n,i,{a-1}}   \right)^{2}.
\label{eqn_temp}  
\end{align}

Due to the  complexity from the nonlinear formulation, the terms in Equation \eqref{eqn:linear_A_prime} and Equation~\eqref{eqn:linear_A} require some computations. In particular, the first term in Equation~\eqref{eqn:linear_A_prime} is rewritten as
\begin{multline}
( {g({\varphi^{n,i-1})}} \bsigma(  \delta \bu^{n,i,a} ), \bfeps(\bfw)) =
\Bigg({g({\varphi^{n,i-1})}}  
\Bigg(
\frac{ 2\mu \left(\dfrac{ \nabla \delta\bu^{n,i,a} + \nabla  \delta\bu^{T\:n,i,a}  }{2}\right)  + \lambda \, (\nabla \cdot  \delta \bu^{n,i,a})\,  \bf{I} }{\left( 1 - \left( \beta  | \mathbb{E}^{1/2} \left[ \bfeps \right]|\right)^{\alpha}\right)^{1/\alpha}}
\\ + 
\frac{\beta^{\alpha}    \theta_{1}\{  \bu^{n,i,a-1} \}  \theta_{2}\{  \bu^{n,i,a-1},  \delta\bu^{n,i,a} \}  \mathbb{E}[ \bfeps ]}{\left( 1 - \beta^{\alpha} | \mathbb{E}^{1/2} \left[  \bfeps \right]|^{\alpha}\right)^{1 + 1/\alpha}} \Bigg) \colon\left(\frac{\nabla \bfw  + \nabla \bfw^{T}}{2}\right) \Bigg),
\end{multline}

where
\begin{align}
\theta_{1}\{ \bu  \} &:= \left| \mathbb{E}^{1/2} \left[  \bfeps   \right] \right|^{\alpha-2},  \\
\theta_{2}\{ \bu,  \delta  \bu  \} &:= \left(   \left| \mathbb{E}^{1/2} \left[  \bfeps   \right] \right| \right)^{\prime}  \notag \\
& = 2\mu \, \left(\dfrac{\nabla   \bu   + \nabla  \bu^{T} }{2}\right) \colon \left(\dfrac{\nabla \delta \bu   + \nabla \delta \bu^{T}}{2}\right)   +  \lambda \, \left( \nabla \cdot   \bu   \right) \left( \nabla \cdot  \delta  \bu    \right). 
\
\end{align}

Moreover, the first term in Equation~\eqref{eqn:linear_A} is derived as 
\begin{multline}
({g({\varphi^{n,i-1})}} \bsigma( \bu^{n,i,a-1} ) , \bfeps(\bfw))  \\
=\Bigg({g({\varphi^{n,i-1})}}  
\Bigg(
\frac{ 2\mu \left(\dfrac{ \nabla  \bu^{n,i,a-1} + \nabla  \bu^{T\:n,i,a-1}}{2}  \right)  + \lambda \, (\nabla \cdot \bu^{n,i,a-1}) \,  \bf{I} }{ \left( 1 - \left( \beta  | \mathbb{E}^{1/2} \left[ \bfeps \right]|\right)^{\alpha}\right)^{1/\alpha}}\Bigg)\colon\left( \frac{\nabla \bfw  + \nabla \bfw^{T}}{2}  \right)\Bigg).
\end{multline}

\subsubsection{Step 2. Solve the phase-field subproblem} 
Secondly, the phase-field subproblem (Equation~\eqref{eqn:main2}) 
is solved with the displacement and phase-field values given from the previous iteration $\{ \bu^{n,i}, \varphi^{n,i-1},~\text{and}~\varphi^{n-1}\}$. 

Given $\bu^{n,i}, \varphi^{n,i-1}$, and $\varphi^{n-1}$, we seek for $\varphi^{n,i}_h \in W_h$ satisfying 
\begin{equation}
A_2(\varphi^{n,i}_h,\psi) = 0, \ \ \forall \psi \in W_h,
\label{eqn:pf} 
\end{equation}
where
\begin{multline}\label{eqn:A_2_lin_nonlin}
A_2( \varphi^{n,i}, \psi)
:=
(1-\kappa)( \varphi^{n,i} \bsigma( \bu^{n,i}): \bfeps( \bu^{n,i}) , \psi )
 -
 G_c (  \dfrac{1}{\xi} ( 1-  \varphi^{n,i})  , \psi) 
 +
 G_c ( {\xi} \nabla  \varphi^{n,i} ,\nabla \psi)  \\
 + (\eta^i (\omega_\gamma^{n,i} + \gamma ( \varphi^{n,i} - \varphi^{n-1})), \psi) 
 +
 L_\varphi( \varphi^{n,i} - \varphi^{n,i-1}  ,\psi).
\end{multline}
{Here the last term is the L-scheme stabilization term with a positive constant value $L_\varphi$, and $\eta^i \in L^\infty(\Lambda)$ is defined as }
$$
\eta^i(x) :=
\begin{cases}
1, \ \ \text{ if }  \omega_\gamma^{n,i} (x) + \gamma (\varphi^{n,i}(x) - \varphi^{n-1}(x))~ {>}~0, \\
0, \ \ \text{ if }  \omega_\gamma^{n,i} (x) + \gamma (\varphi^{n,i}(x) - \varphi^{n-1}(x))~ {\leq}~0, \\
\end{cases}
$$
to replace the operator $[\cdot]^+$.

To solve the nonlinear problem of Equation~\eqref{eqn:pf}, we employ the Newton iteration algorithm coupled with an appropriate line search. 
Thus, we find {$\delta \varphi^{n,i,b} \in W_h$} by solving 
\begin{equation}
A^{\prime}_2(\varphi^{n,i,{b-1}})(\delta \varphi^{n,i,b}, \psi) = -A_2(\varphi^{n,i,{b-1}})(\psi), \ \ \forall \psi \in V_h,
\end{equation}
for the iterations step, $b=1,2,\cdots,$ 
until $\| \delta \varphi^{n,i,b}\| \leq \varepsilon_b$. 
Then we update 
\begin{equation}
\varphi^{n,i,b} = \varphi^{n,i,b-1} + \omega_\varphi \, \delta \varphi^{n,i,b},
\end{equation}
in which $\omega_{\varphi}$ is a line search parameter and $\omega_{\varphi} \in [0,  1]$.
Here the Jacobian of $A_2(\varphi(\psi))$ applied to a direction of 
$\delta \varphi$ is 
\textcolor{black}{\begin{multline}
A^{\prime}_2(\varphi^{n,i,{b-1}})( \delta \varphi^{n,i,b}, \psi) := 
(1-{\kappa})( \delta \varphi^{n,i,b}~\bsigma(\bu^{n,i}): \bfeps(\bu^{n,i}) , \psi )
 +
 G_c ( \dfrac{1}{\xi}  \, \delta \varphi^{n,i,b}  , \psi) \\
 +
 G_c ( \xi \, \nabla \delta \varphi^{n,i,b}, \nabla \psi) 
 + \eta^i \gamma( \delta \varphi^{n,i,b} , \psi) 
 +
 L_\varphi( \delta \varphi^{n,i,b} ,\psi),
\end{multline} }
and 
\begin{multline}
A_2(  \varphi^{n,i,{b-1}})(\psi)
:=
(1-{\kappa})(  \varphi^{n,i,b-1}~\bsigma(\bu^{n,i}): \bfeps(\bu^{n,i}) , \psi )
 -
 G_c (  \dfrac{1}{\xi} \, ( 1- \varphi^{n,i,b-1})  , \psi)  \\
 +
 G_c ( {\xi} \, \nabla \varphi^{n,i,b-1} ,\nabla \psi) 
 + (\eta^i (\omega_\gamma^{n,i} + \gamma ( \varphi^{n,i,b-1} - \varphi^{n-1} )), \psi) 
 +
 L_\varphi( \varphi^{n,i,b-1} - \varphi^{n,i-1}  ,\psi).
\end{multline}   
If the Newton iteration converges, we set 
$$\varphi^{n,i} = \varphi^{n,i,b}.$$
We note that the choice of the $\bsigma$ is either Equation \eqref{eqn:sig_lin} for the linear case, 
or Equation \eqref{eq:nonlinear-sigma-plus} for the nonlinear strain-limiting case,
depending on the mechanics subproblem that we solve.

As we discussed in the previous section, the augmented-Lagrangian iteration is embedded in the L-scheme iteration. Thus, the augmented term 
$\omega_\gamma^{n,i}$ is updated every staggered step of $i$:
\textcolor{black}{\begin{equation}
\omega_\gamma^{n,i} = [ \omega_\gamma^{n,i-1} + \gamma (  \varphi^{n,i,b-1} - \varphi^{n-1} ) ]^+.
\end{equation}}
We also note that the phase-field function has three different categories for the iteration index: the previous timestep index $n-1$, the staggered step of the L-scheme iteration index $i~\text{and}~i-1$, and the Newton iteration index 
$b~\text{and}~b-1$. 
Whereas, displacement value is given as $\bu^{n,i}$, which is computed from the first step of the L-scheme.  

{Finally,  we employ both mechanics subproblem residual 
$\|A_1(\bu^{n,i},\bfw)\|\leq \text{T}_{\text{OL}}$
and
phase-field subproblem residual  
$\|A_2(\varphi^{n,i}_h,\psi)\| \leq \text{T}_{\text{OL}}$ 
as the stopping criteria for both L-scheme and augmented Lagrangian. 
If the whole iteration converges, we obtain
$$
 \bu^{n} = \bu^{n,i,a}  \ \ \  \text{ and } \ \ \ 
\varphi^{n} = \varphi^{n,i,b}.
$$
}
\section{Numerical Examples}
\label{sec:examples}

In this final section, we present several  numerical examples to verify and validate the  proposed nonlinear algorithm.
Moreover, we illustrate the capabilities and the effectiveness of the framework. 
The code is based on the open-source finite element package deal.II \cite{dealII91} and all the computations are performed utilizing high performance computing machines at Texas A\&M University - Corpus Christi. 
For the nonlinear strain-limiting (NLSL) model, the computations are developed by the authors based on the previous studies \cite{brun2019iterative,wheeler2014augmented}.  

From the displacement ($\bu$) obtained from the governing equations coupled with the phase-field, i.e., Equation~\eqref{eq:blaws} and 
Equation~\eqref{eqn:linear_pf_new}, respectively, the stress values are calculated using Hooke's law (Equation~\eqref{eq:HLaw}) for both models. Each strain value calculation is based on each model: $\bfeps$ from Equation~\eqref{LinStrain} for LEFM, and $\bfeps_{{\text{NL}}}$ for 
 NLSL with Equation~\eqref{eqn:nonIinear1}. 
 
\subsection{Example 1: The error convergence tests}
In the first example, the error convergence is tested to verify the implementation for 
NLSL formulation presented in the previous sections. 
For simplicity, only the mechanics subproblem is considered by neglecting the phase-field variable. 
Thus, we set the phase-field to be a constant one for the whole domain ($\varphi = 1$) and ${\kappa=0}$. 

\begin{table}[!h]
\centering
\begin{tabular}{|c|l||l|l||l|l|}
\hline
\multirow{2}{*}{Cycle} & \multicolumn{1}{c||}{\multirow{2}{*}{h}} & \multicolumn{2}{c||}{Linear}                               & \multicolumn{2}{c|}{Nonlinear}                            \\ \cline{3-6} 
                       & \multicolumn{1}{c||}{}                   & \multicolumn{1}{c|}{L2 Error} & \multicolumn{1}{c||}{Rate} & \multicolumn{1}{c|}{L2 Error} & \multicolumn{1}{c|}{Rate} \\ \hline\hline
1                      & 0.25                                    & 0.033493958414                & 0.0                       & 0.031402524561                & 0.0                       \\ \hline
2                      & 0.125                                   & 0.008457780816                & 2.6942                    & 0.007450392935                & 2.8163                    \\ \hline
3                      & 0.0625                                  & 0.002119761659                & 2.3542                    & 0.001790875453                & 2.4253                    \\ \hline
4                      & 0.03125                                 & 0.000530273421                & 2.1788                    & 0.000437507028                & 2.2160                    \\ \hline
5                      & 0.015625                                & 0.000132589164                & 2.0898                    & 0.000108024578                & 2.1088                    \\ \hline
6                      & 0.0078125                               & 0.000033148594                & 2.0450                    & 0.000026842623                & 2.0540                    \\ \hline
\end{tabular}
\caption{Example 1. The results of $L^2$ error convergence test of the approximated displacement for the linear {(LEFM)} and the nonlinear {(NLSL)}  mechanics subproblem  are illustrated. We observe the optimal convergence for both cases.}
\label{fig:ex1}
\end{table}

Here, the given exact solution for the mechanics subproblem is defined as 
\begin{equation}\label{eq:Ex0}
\bu(x,y) :=(\sin{x}\sin{y},\cos{x}\cos{y}),
\end{equation}
in the computational domain $\Lambda = [0,1]^2$.
The right hand side and the boundary conditions are chosen accordingly to satisfy the homogeneous boundary conditions on $\partial \Lambda$. 
In addition, {Lam\'{e}} coefficients are set as $\lambda=\mu=0.01$ and 
the nonlinear parameters are given as {\color{black}$(\alpha, \beta)=(0.1,0.1)$}.
Six computations on uniform meshes were computed where the mesh size $h$ is divided by two for each cycle, and the corresponding number of cells for each cycle is $4, 16, 64, 256, 1024$, and $4096$. 

The {results} of the $L^2(\Lambda)$ errors for the approximated displacement solution versus the mesh size $h$ are shown in 
Table~\ref{fig:ex1}. We observed the expected optimal convergence rate for both linear and nonlinear cases for our mechanics subproblem.

\subsection{Example 2: Strain-limiting effects for a static fracture}
In this example, we compare the presented~{NLSL} model with 
 {LEFM} model in the domain with a static fracture. 
 In $\Lambda = [0,1]^2$, the initial fracture is described as a slit on $(0.5,0.5)-(1.0,0.5)$.  
The Dirichlet boundary condition $\bu = (0,\bar{u}_\textsc{top})$ is employed at the top of the boundary, $\Gamma_{D_1}$, where the values of 
$\bar{u}_\textsc{top}$ are chosen differently with respect to the test cases. 
On  the bottom of the boundary,  $\Gamma_{D_2}$, 
only the y-component is imposed with zero value but the x-component is traction-free. 
The homogeneous traction-free Neumann boundary condition is employed for the left and right boundaries, $\Gamma_N$, including the slit.
See Figure \ref{Fig:ex2_setup} for more details. 
The initial mesh is refined 7 times globally, thus $h = 0.0078125 $.
Moreover, 
we utilized the linear problem for the initial guess for the first nonlinear Newton 
iteration {of NLSL} to expedite the convergence. 
\begin{figure}[!h]
\centering
\begin{minipage}{0.45\textwidth}
\centering
\begin{tikzpicture}
\draw (0,0) -- (3,0) -- (3,3) -- (0,3) -- (0,0);

\draw [line width=0.5mm,  blue]  (1.5,1.5) -- (3,1.5);

\node at (-0.3,-0.25)   {$(0,0)$};
\node at (3.3,3.2)   {$(1,1)$};

\node at (-0.3, 1.5)   {$\Gamma_N$};
\node at (3.3, 1.5)   {$\Gamma_N$};
\node at (1.5, -0.25)   {$\Gamma_{D_2}$};
\node at (1.5, 3.2)   {$\Gamma_{D_1}$};

\draw[->] (0.2,3.1) -- (0.2,3.5);
\draw[->] (1.,3.1) -- (1.,3.5);
\draw[->] (2.,3.1) -- (2.,3.5);
\draw[->] (2.8,3.1) -- (2.8,3.5);

\node at (1.5,3.7) {$\bar{u}_\textsc{top}$};
\end{tikzpicture}
\caption{Example 2. A setup and the boundary conditions: the blue line indicates the slit and the arrows on the top denote the axial traction.}
\label{Fig:ex2_setup}
\end{minipage}
\begin{minipage}{0.45\textwidth}
\footnotesize
\centering
\begin{tabular}{c|c||ccl}
& & $\bar{u}_\textsc{top}$ & $\beta$ & $\alpha$   \\
\hline
\multirow{4}{*}{\texttt{CASE 1}} & i   & \multirow{4}{*}{2.0} & \multirow{4}{*}{0.04} & 2    \\
                  & ii  &                    &                    & 1    \\
                  & iii &                    &                    & 0.5  \\
                  & iv  &                    &                    & 0.25 \\
\hline    				
\multirow{4}{*}{\texttt{CASE 2}} & i   & \multirow{4}{*}{1.0} & \multirow{4}{*}{0.09} & 2    \\
                  & ii  &                    &                    & 1    \\
                  & iii &                    &                    & 0.5  \\
                  & iv  &                    &                    & 0.25 \\
\hline    				
\multirow{4}{*}{\texttt{CASE 3}} & i   & \multirow{4}{*}{0.5} & \multirow{4}{*}{0.18} & 2    \\
                  & ii  &                    &                    & 1    \\
                  & iii &                    &                    & 0.5  \\
                  & iv  &                    &                    & 0.25 \\
\hline    				
\multirow{4}{*}{\texttt{CASE 4}} & i   & \multirow{4}{*}{0.1} & \multirow{4}{*}{0.92} & 2    \\
                  & ii  &                    &                    & 1    \\
                  & iii &                    &                    & 0.5  \\
                  & iv  &                    &                    & 0.25 \\
\end{tabular}
\caption{Example 2. Different test cases for $\alpha$ and $\beta$}
\label{tab004}
\end{minipage}
\end{figure}

Here, we test four different cases for the displacement values on the top boundary as $\bar{u}_\textsc{top}=2.0,~1.0,~0.5,~\text{and}~0.1$, corresponding to \texttt{CASE 1, CASE 2, CASE 3}, and \texttt{CASE 4}, respectively. 
As we discussed in Remark \ref{rmk001}, the suitable nonlinear parameter pair of $(\alpha,\beta)$ should be chosen to satisfy the condition of Equation~\eqref{eqn:beta_condition}.
More precisely, we obtain
\begin{equation}\label{eq:Ex002-beta}
0\leq\beta < \left(\frac{1}{ | \mathbb{E}^{1/2} \left[ \bfeps \right]|^{\alpha}}\right)^{1/\alpha},
\end{equation}
and  the condition simplifies to 
$\beta| \mathbb{E}^{1/2} \left[ \bfeps \right]|<1$ by assuming $\alpha  >0$.
In this case, we only need to satisfy $0\leq\beta| \mathbb{E}^{1/2} \left[ \bfeps \right]|<1$ for a given value of $\beta$. 
Therefore, the maximum $\beta$ values (rounding to 2 decimal places) for each cases are presented in Figure~\ref{tab004} for each top boundary condition, $\bar{u}_\textsc{top}$, satisfying the condition in the inequality above with any given positive $\alpha$ 
and by setting {Lam\'{e}} coefficients as $\lambda=\mu=1.0$,

Moreover, to investigate the effects of the parameters ($\alpha, \beta$) for 
NLSL, we vary the choice for the $\alpha$ values. 
Starting with $\alpha=2$,
we arbitrarily set $\alpha$ by reducing in half as shown in  Figure~\ref{tab004}.
Thus, we study a total of 16 different cases for  
NLSL model, and 4 different cases for LFEM (which are identical to the corresponding NLSL models when $\beta=0$)
are also computed for the comparison. 
Eventually, we aim to 
see the maximized strain-limiting effect from the optimized combinations of  ($\alpha, \beta$).

\begin{figure}[!h]
\centering
\includegraphics[width=1.0\textwidth]{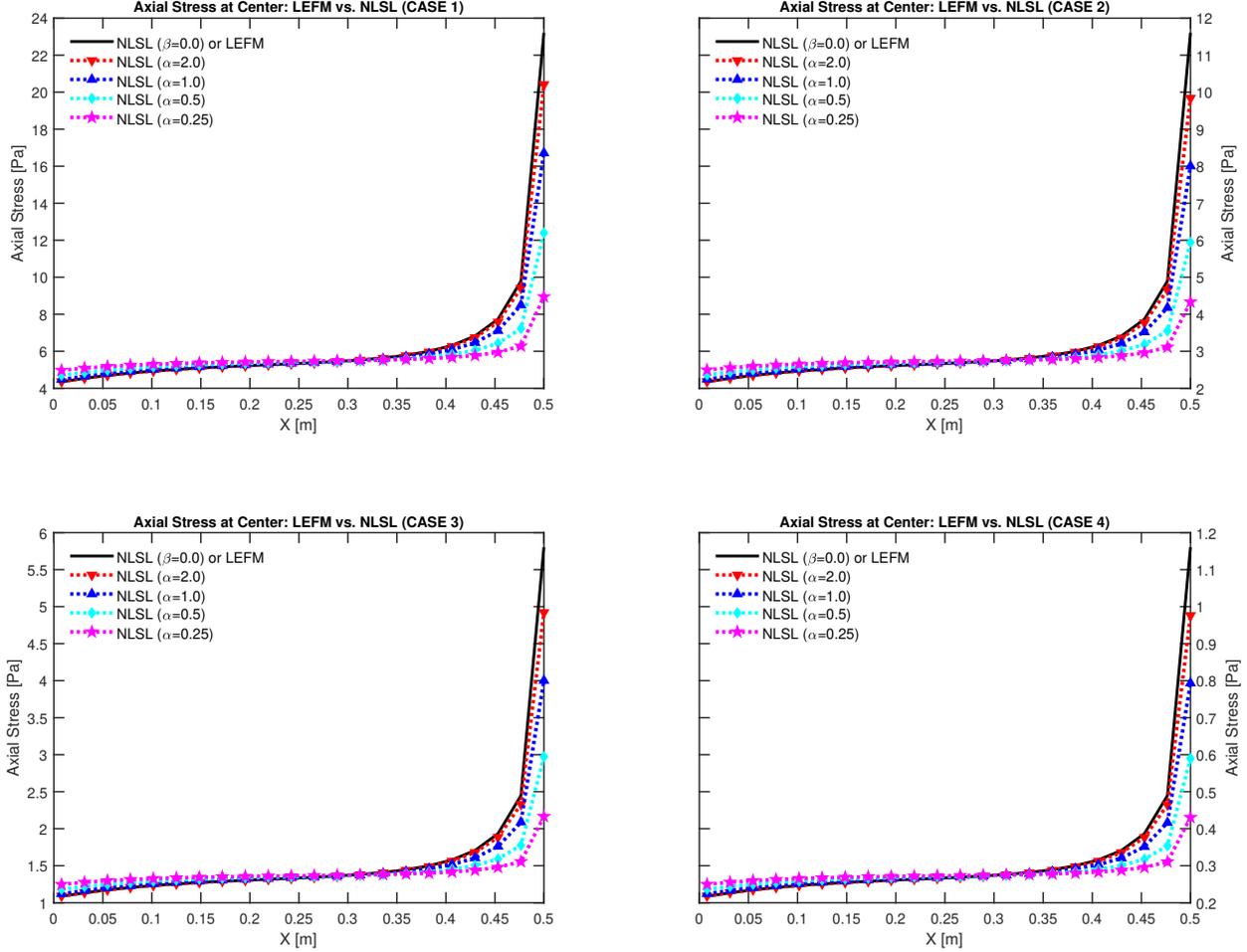}
\caption{{Example 2. Axial stress for each case (from \texttt{CASE 1} in the top left to \texttt{CASE 4} in the bottom right): approaching to the crack-tip, $\text{X}=0.5$, similar singular patterns for the stress values
are shown in each case.}} 
\label{Fig_Ex2_Stress}
\end{figure}

\begin{figure}[!h]
\centering
\includegraphics[width=1.0\textwidth]{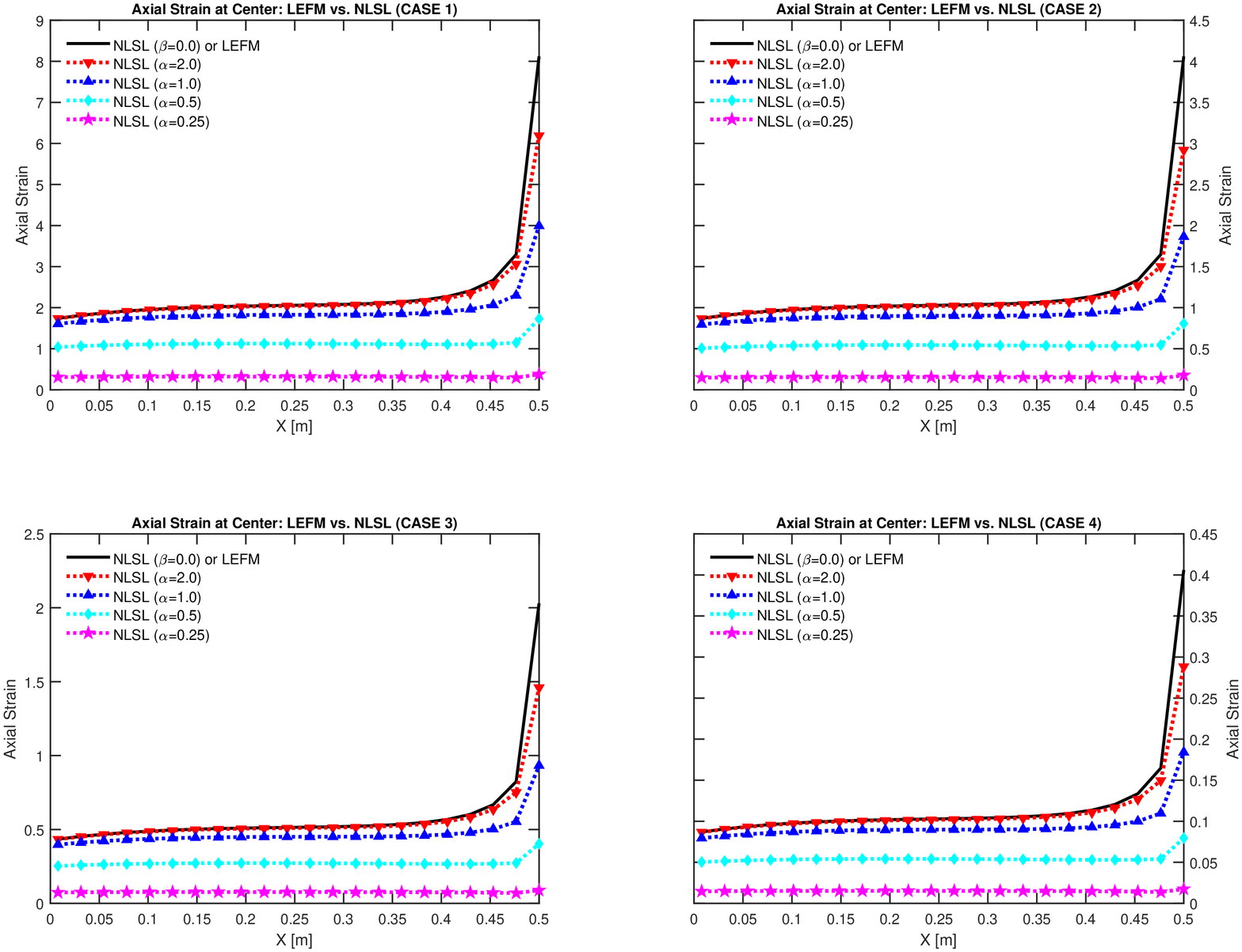}
\caption{Example 2. Axial strain for each case {(from \texttt{CASE 1} in the top left to \texttt{CASE 4} in the bottom right)}: with smaller values of $\alpha$, strain is distinctively limited for the {nonlinear} strain-limiting {(NLSL)} model in each case.}
\label{Fig_Ex2_Strain}
\end{figure}

To this end, we calculate the axial stress and strain along the center line, {$(0, 0.5) - (0.5, 0.5)$}, 
i.e., starting from the left boundary to the starting location of slit. 
The axial stress ($\sigma_{22}$) corresponds to the component of $\bsigma$ {from Hooke's law (Equation~\eqref{eq:HLaw} or $\mathbb{E}(\bfeps)$~}in Equation~\eqref{eqn:stress}), 
whereas the {axial} strain ($\epsilon_{22}$) is calculated {with the corresponding component of $\bfeps$ or $\bfeps_\text{{NL}}$.~}
If we have $\beta=0$ or $\alpha\rightarrow\infty$, then
$\bfeps_\text{{NL}}$ is identical to 
$\bfeps$ for LEFM, without any strain-limiting effect. Also note that we compute the average values of $\sigma_{22}$ and $\epsilon_{22}$ in quadrature points for each cell.

First, Figure~\ref{Fig_Ex2_Stress} illustrates the axial stress values for each case by varying the $\alpha$ and $\beta$ values as shown in Figure~\ref{tab004}. For each case, the stress values are compared with {LEFM, which is identical when $\beta=0$ for {NLSL}}. 
The overall pattern of singular stress value near the slit tip are almost identical {even} with different $\alpha$ and $\beta$ values in each case. 

On the other hand, Figure~\ref{Fig_Ex2_Strain} presents the nonlinear strain-limiting effects of NLSL.~
Here, the axial strain values for each case are illustrated. 
We note that the obvious strain-limiting effect is shown by comparing with the values from {LEFM}. 
The different effects are observed by different choice of the $\alpha$ values. 
With this setup, the most strain-limiting effect occurs with the smallest value of $\alpha=0.25$ given for each case. This is a consistent result from the theory that NLSL becomes LEFM if $\alpha\rightarrow \infty$. 

Finally, from this example, we observe that the nonlinear effects are sensitive to the choice of nonlinear parameters. 
A wise selection of the parameters for maximizing (or optimizing) the strain-limiting effects is necessary \cite{bonito2018finite,fu2019generalized,fu2019constraint}. In addition, 
for the larger strain-limiting effects (i.e., for larger nonlinear effects), more iterations for convergence are required in Newton 
 method.

\subsection{Example 3: A {static} phase-field fracture} 
In this example, we replace the fracture representation in Example 2 with the phase-field approach and investigate NLSL model.
~Most of the setup is the same as the previous example, but here the phase-field variable $\varphi$ is employed to describe the fracture.
Thus, in the computational domain $\Lambda = [0,1]^2$, 
a (prescribed) initial crack with length $l_0 = {0.5}$ is placed 
on $(0.5,1.)\times(0.5-h_{\min}, 0.5+h_{\min}) \subset \Lambda $. 
The initial phase-field values are set to zero for the initial fracture
described above and $\varphi = 1$ otherwise. This replaces the slit in the previous example.

\begin{figure}[!h]
\centering
\includegraphics[width=0.355\textwidth]{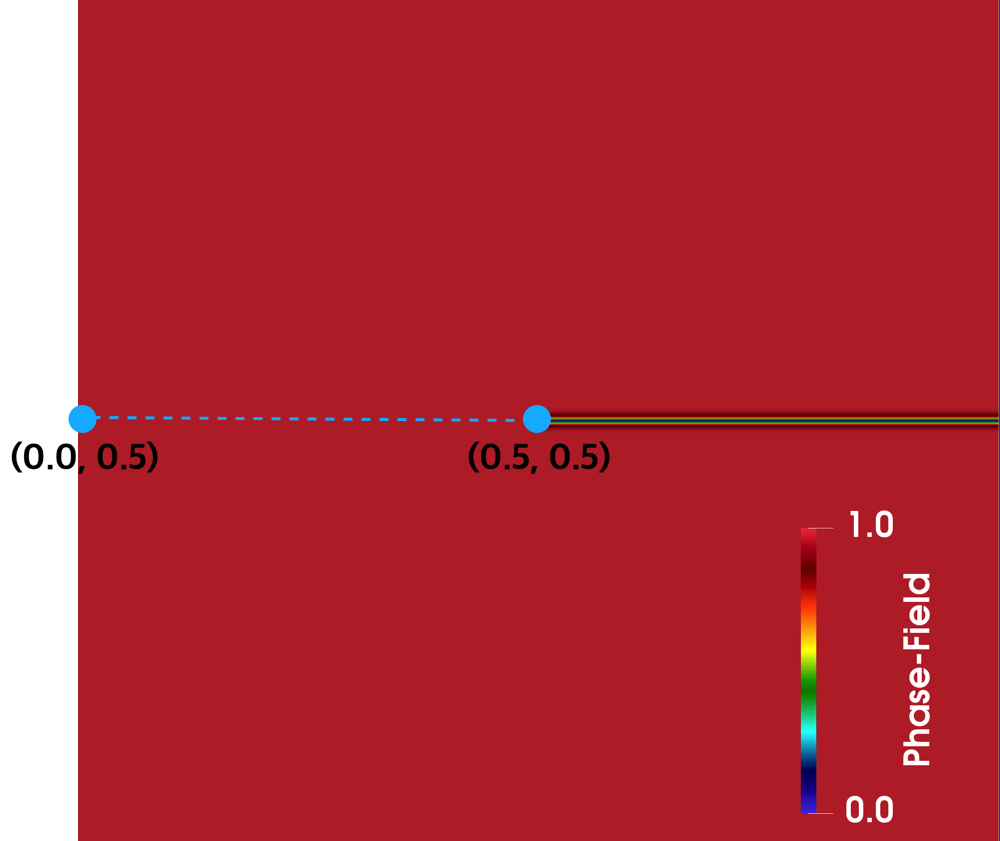}
\includegraphics[width=0.3275\textwidth]{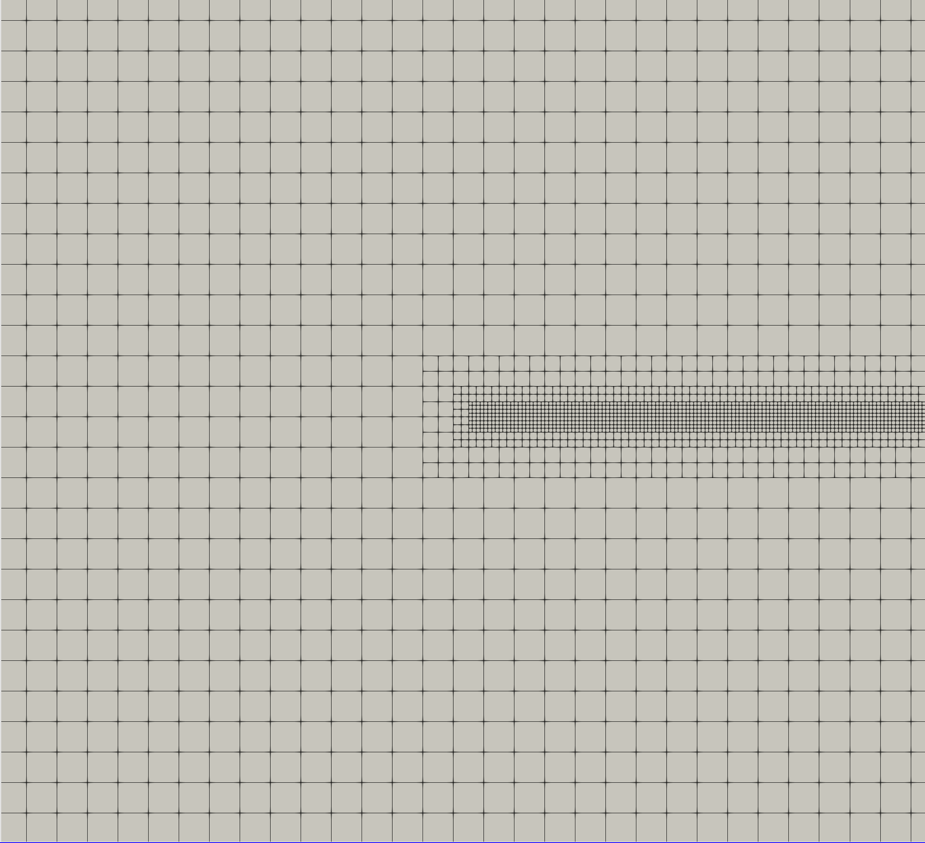}
\caption{{Example 3. (Left) illustrates the setup with an initial phase-field fracture. As the previous example, the stress and strain values are plotted on the dashed line (0.0, 0.5)-(0.5,0.5).  (Right) adaptive mesh refinement is employed near the fracture.}}
\label{fig:ex3_setup}
\end{figure}

The initial mesh is seven times uniformly refined as the previous example but here  three additional levels of adaptive mesh refinement is employed near the fracture, where $\varphi<0.9$, 
resulting in $h_{\min}=0.0009765625$. 
For the phase-field, homogeneous Neumann condition is employed and
the regularization parameters are chosen as $\xi = 2h_{\min}$, and $\kappa = \num{e-10}h_{\min}$.
See Figure \ref{fig:ex3_setup} for more details. 

The same displacement boundary conditions on $\Gamma_{D_1}$ and $\Gamma_{D_2}$ as the previous example are employed, 
but here we set $\bar{u}_\textsc{top}=0.0001$. 
For the coupling between the mechanics and the phase-field, the presented L-scheme is utilized by choosing the L-constant as 
$10^{-6}$ for both mechanics and phase-field (i.e., $\text{L}_u=\text{L}_\phi=1e$-6). 
The stopping criteria  for the staggered L-scheme is 
$\text{T}_{\text{OL}}=10^{-6}$,
and the stopping criteria for the newton method for both displacement  and phase-field  are set as $\varepsilon_a = \varepsilon_b = 10^{-8}$. 
Note that {for mechnanics subproblem in NLSL,} only the first Newton iteration 
is utilizing the initial guess from the linear problem for faster convergence. 
In addition, the penalty parameter $\gamma=10^4$ is set for the irreversibility condition. 
The critical energy release rate is chosen as 
$G_c = \SI{5}{\newton\per\metre}$.  
Then, all the other numerical and physical parameters are the same as the previous example.

With the given conditions above, here we investigate the effects of the nonlinear parameters for both $(\alpha,\beta)$.
First, by Equation~\eqref{eq:Ex002-beta},  we obtained the maximum of $\beta$ as $\beta_{\max}=127$ for varying $\alpha$.
In addition, we varied the choice for the value of $\beta < \beta_{\max}$ by fixing the value of $\alpha$. 
Thus, as shown in Table~\ref{tab005}, we investigated the total 6 different cases  for  NLSL model.
\begin{table}[!h]
\centering
\begin{tabular}{c|c||c|l||c|l}
& & \footnotesize Fixed Parameter & \footnotesize Value & \footnotesize Changing Parameter & \footnotesize Value   \\
\hline
\multirow{3}{*}{\footnotesize\texttt{CASE 1}} & i   & \multirow{3}{*}{$\beta$} & \multirow{3}{*}{127} & \multirow{3}{*}{$\alpha$}   & 2\\
                  & ii  &                    &                    &     &1\\
                  & iii  &                    &                    &  &0.5\\
\hline    				
\multirow{3}{*}{\footnotesize\texttt{CASE 2}} & i   & \multirow{3}{*}{$\alpha$} & \multirow{3}{*}{0.25} & \multirow{3}{*}{$\beta$} & 1    \\
                  & ii  &                    &                    &     &10\\
                  & iii &                    &                    &   &50\\
\end{tabular}
\caption{Example 3. Different test cases for $\alpha$ and $\beta$.}
\label{tab005}
\end{table}

\begin{figure}[!h]
\centering 									   
\includegraphics[width=1.0\textwidth]{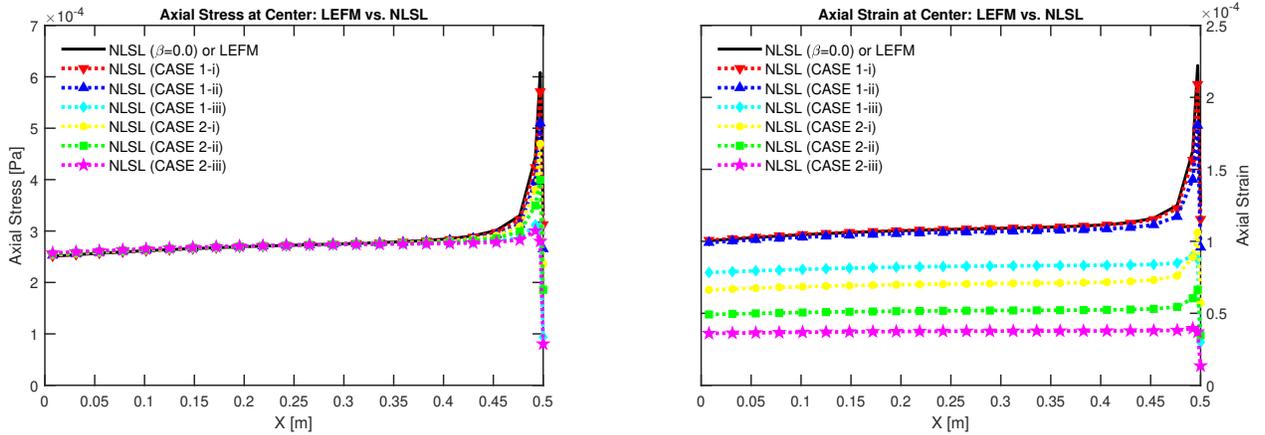}
\caption{Example 3. Axial stress (Left) and strain (Right) values for each case. We observe the strain-limiting effect near the tip of phase-field fracture with an appropriate choice of $\alpha$ and $\beta$. }
\label{fig:ex3_results}
\end{figure}

Figure~\ref{fig:ex3_results} presents the effect of our proposed nonlinear strain-limiting model with the phase-field approach. 
Here, the axial stress ($\sigma_{22}$) and strain ($\epsilon_{22}$) values along the center line {$(0, 0.5) - (0.5, 0.5)$} are computed for both {LEFM and NLSL} as the previous example. 
Overall, the strain-limiting effect is well presented through each combination of $(\alpha, \beta)$ with the phase-field fracture. 
Especially, we observe the dramatic limiting effect of strain when $\alpha<1.0$.

We note that  
 the values of stress and strain are reduced near the crack-tip region (Figure~\ref{fig:ex3_results}),
 due to the phase-field, since there is no mechanics when $\varphi = 0$. In particular, the stress with the phase-field is defined as $\bsigma_{\varphi}:={g(\varphi)\bfsig} = ((1-\kappa)\varphi^2 + \kappa) \,  \bfsig$, and the stress values approach to zero 
near the front of the phase-field crack-tip.

\subsection{Example 4: A quasi-static propagating fracture} 
In this final example, we consider the fracture propagation by employing the quasi-static phase-field approach with the given boundary condition for each timestep. 
The basic setup including the initial and boundary conditions is similar to Example 2 (See Figure~\ref{Fig:ex2_setup}), but in this example, we march the timesteps to propagate the given fracture.
The timestep size is chosen as $\Delta t = {0.0001}$ 
and we set 
$\bar{u}_\textsc{top}= t$,  thus the displacement imposed at the top boundary is increased by marching the timesteps.
{The total number of timesteps is set to N=$50$, which is enough to observe the full propagation of the fracture. The initial mesh is refined 7 times globally and we pre-refine around the expected crack path 
($0.0 \leq x \leq 0.6$  $0.4<y<0.6$)  locally for two more levels. 
Here, $h_{\min} = 0.002$.}

For NLSL model, we set the nonlinear parameter pair as 
$(\alpha,\beta)=(0.25, 4.8\times10^{-4})$ to satisfy the condition Equation~\eqref{eq:Ex002-beta}. 
Since the displacement load is increased in every timestep, we take the minimum of $\beta$ for the entire timesteps.
In this case, we note that the condition from Equation~\eqref{eq:Ex002-beta} is enforced 
throughout the whole simulation.
The same Newton iteration tolerance and  staggered L-scheme coefficients as the previous example is chosen, and the penalty parameter for the irreversibility condition is set as {$\gamma=$1e-7. }

\begin{figure}[!h]
\centering
\subfloat[$n=35$]{\includegraphics[width= 0.3\textwidth]{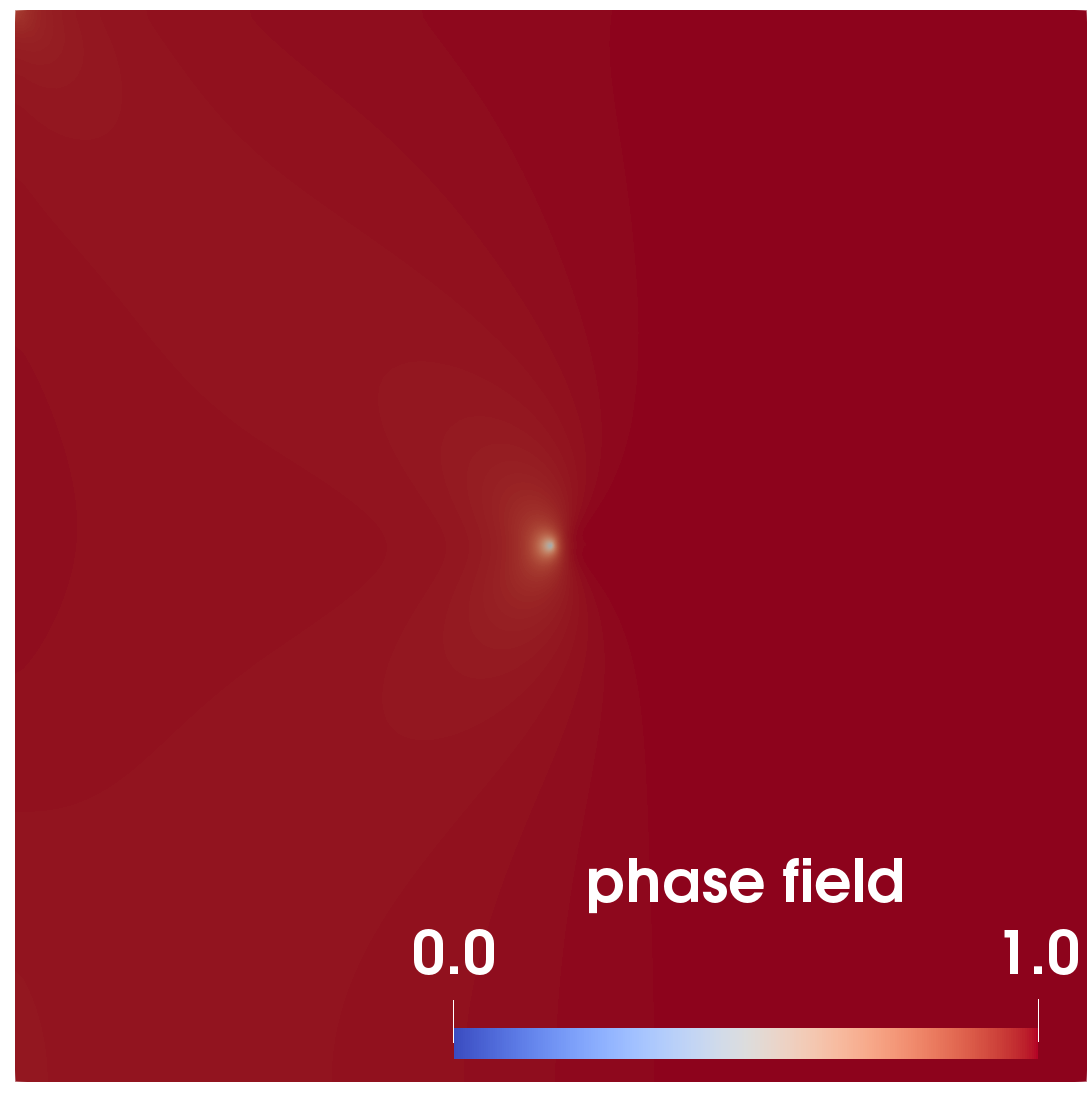}}
\hspace*{0.2in}
\subfloat[$n=35$]{\includegraphics[width= 0.3\textwidth]{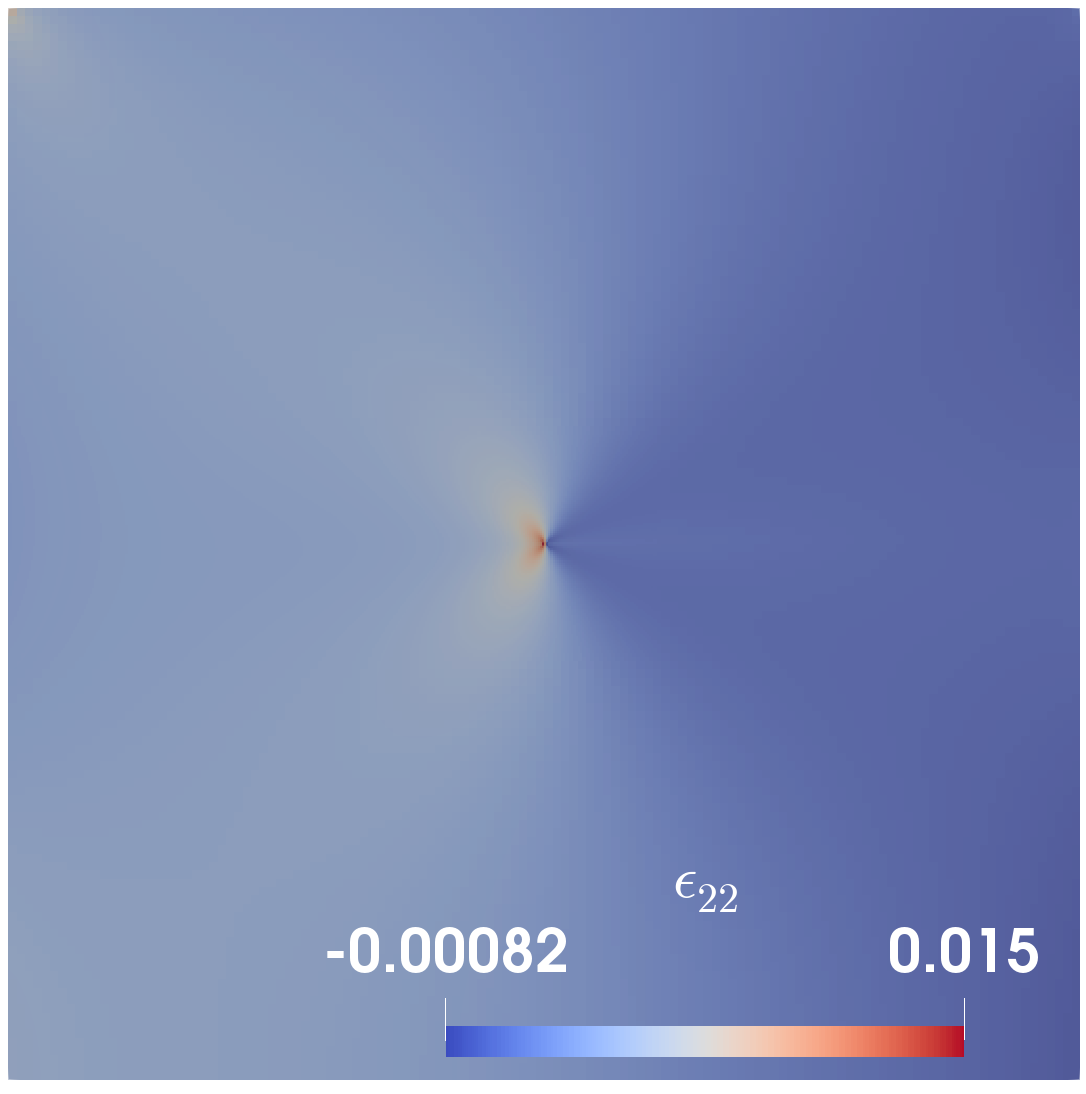}} \\
\subfloat[$n=38$]{\includegraphics[width= 0.3\textwidth]{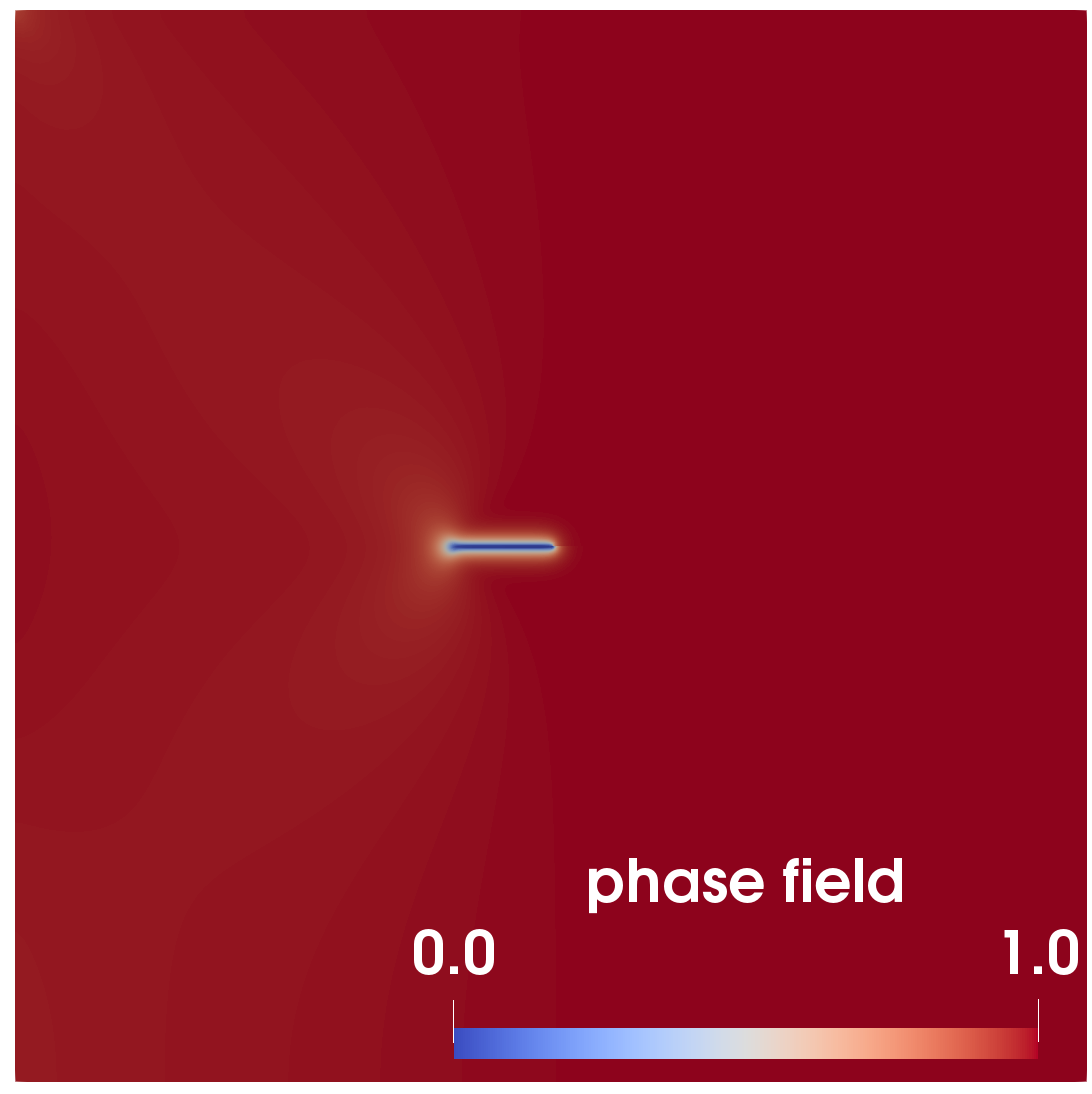}}
\hspace*{0.2in}
\subfloat[$n=38$]{\includegraphics[width= 0.3\textwidth]{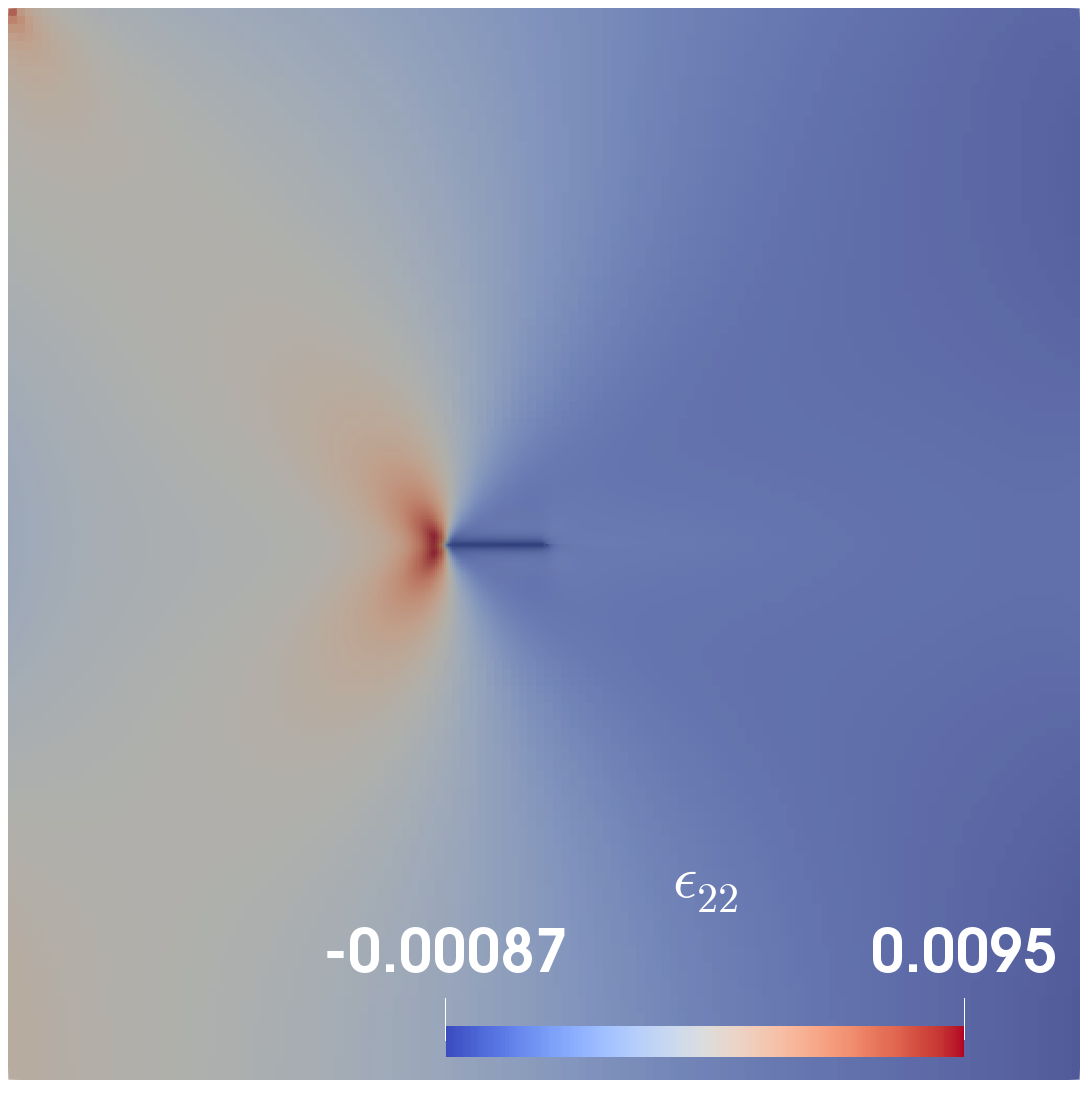}} \\
\subfloat[$n=42$]{\includegraphics[width= 0.3\textwidth]{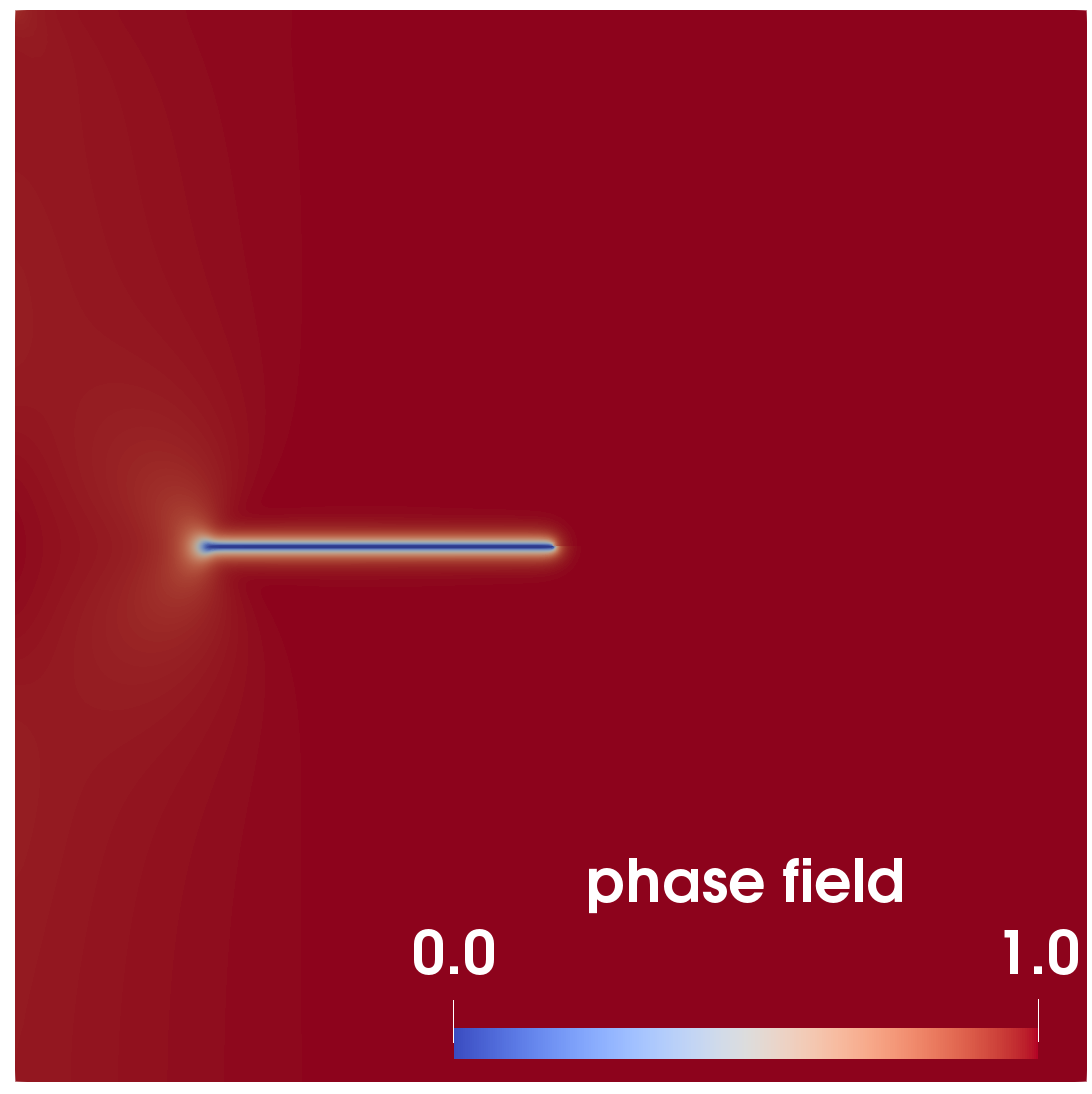}}
\hspace*{0.2in}
\subfloat[$n=42$]{\includegraphics[width= 0.3\textwidth]{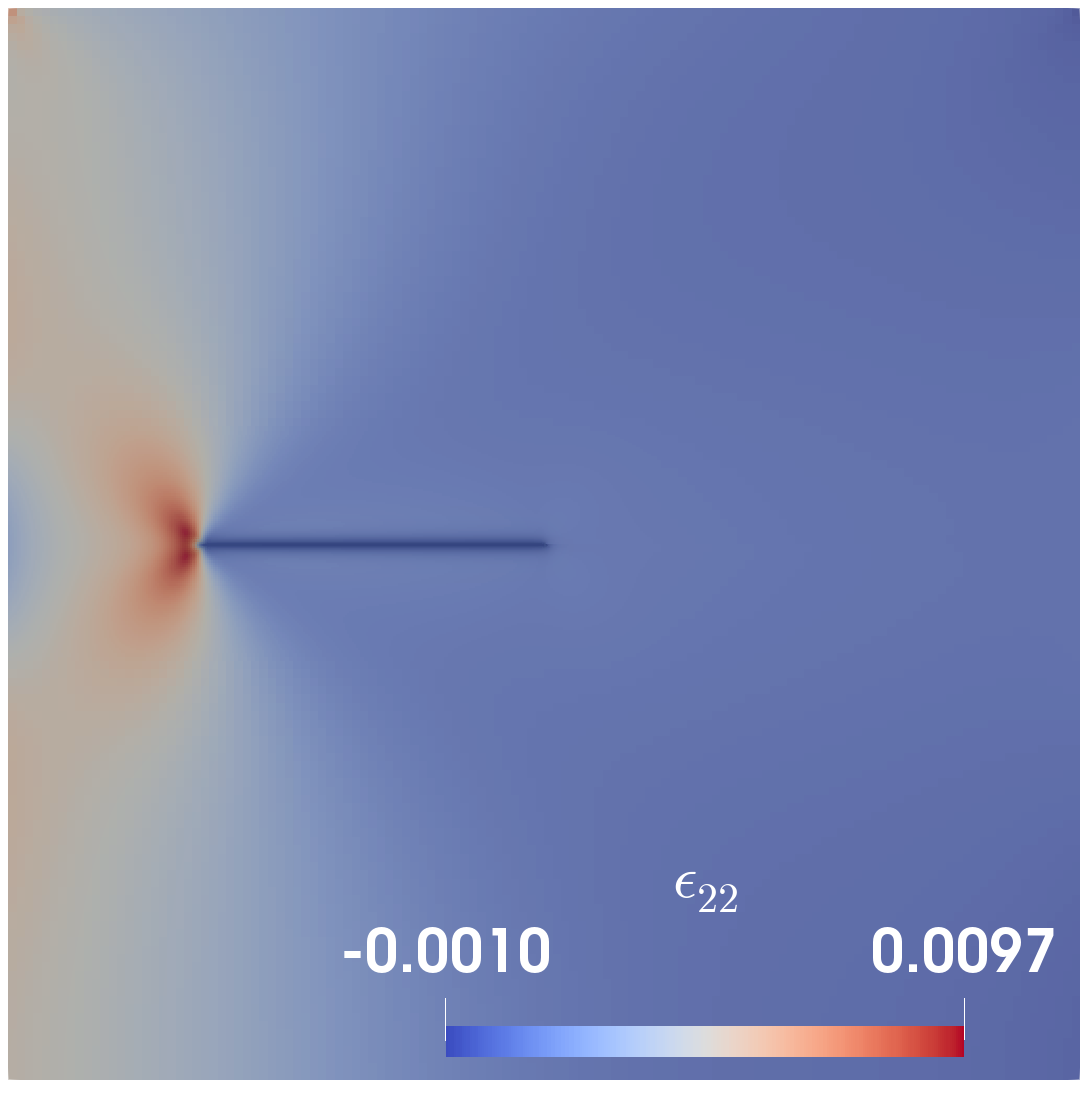}} 
\caption{ {Example 4. (Left) illustrate the phase-field values during crack evolution  for each timestep with LEFM. 
(Right) present the corresponding $\epsilon_{22}$ values for each case. 
The dark blue line indicates the corresponding fracture (phase-field) from the left figure.}}
\label{fig:ex4_linear}
\end{figure}

\begin{figure}[!h]
\centering
\subfloat[$n=25$]{\includegraphics[width= 0.3\textwidth]{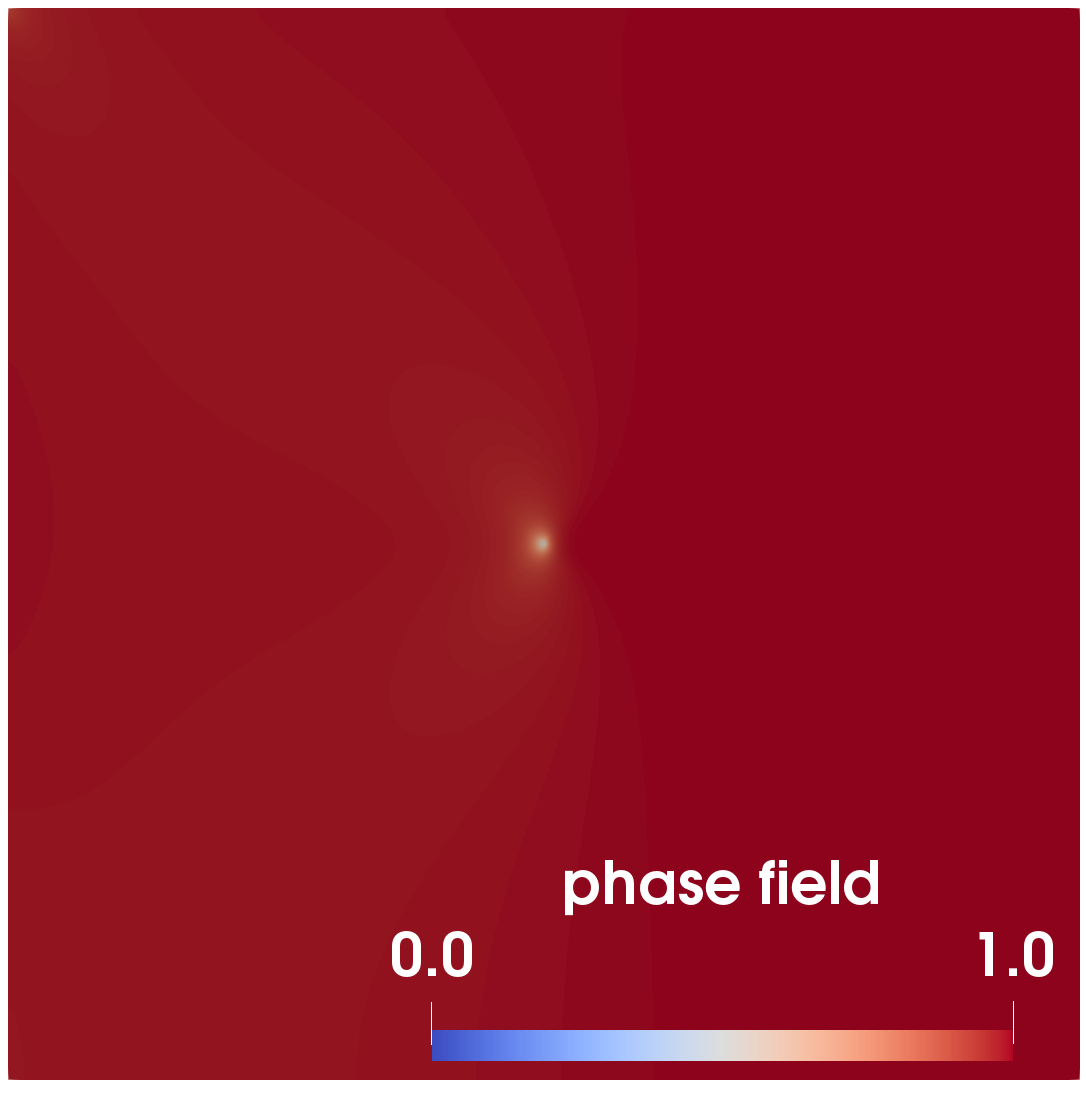}}
\hspace*{0.2in}
\subfloat[$n=25$]{\includegraphics[width= 0.3\textwidth]{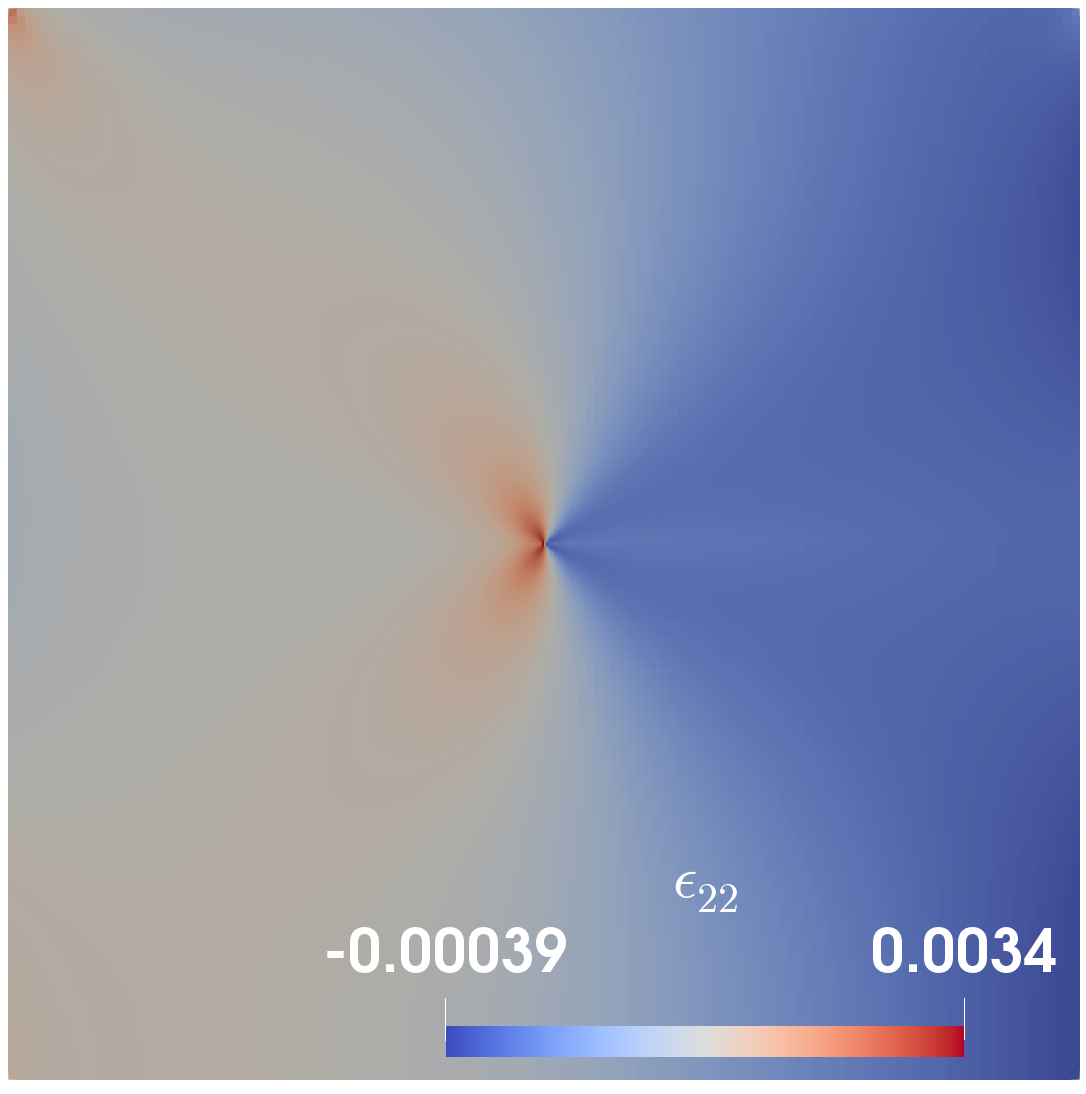}} \\
\subfloat[$n=30$]{\includegraphics[width= 0.3\textwidth]{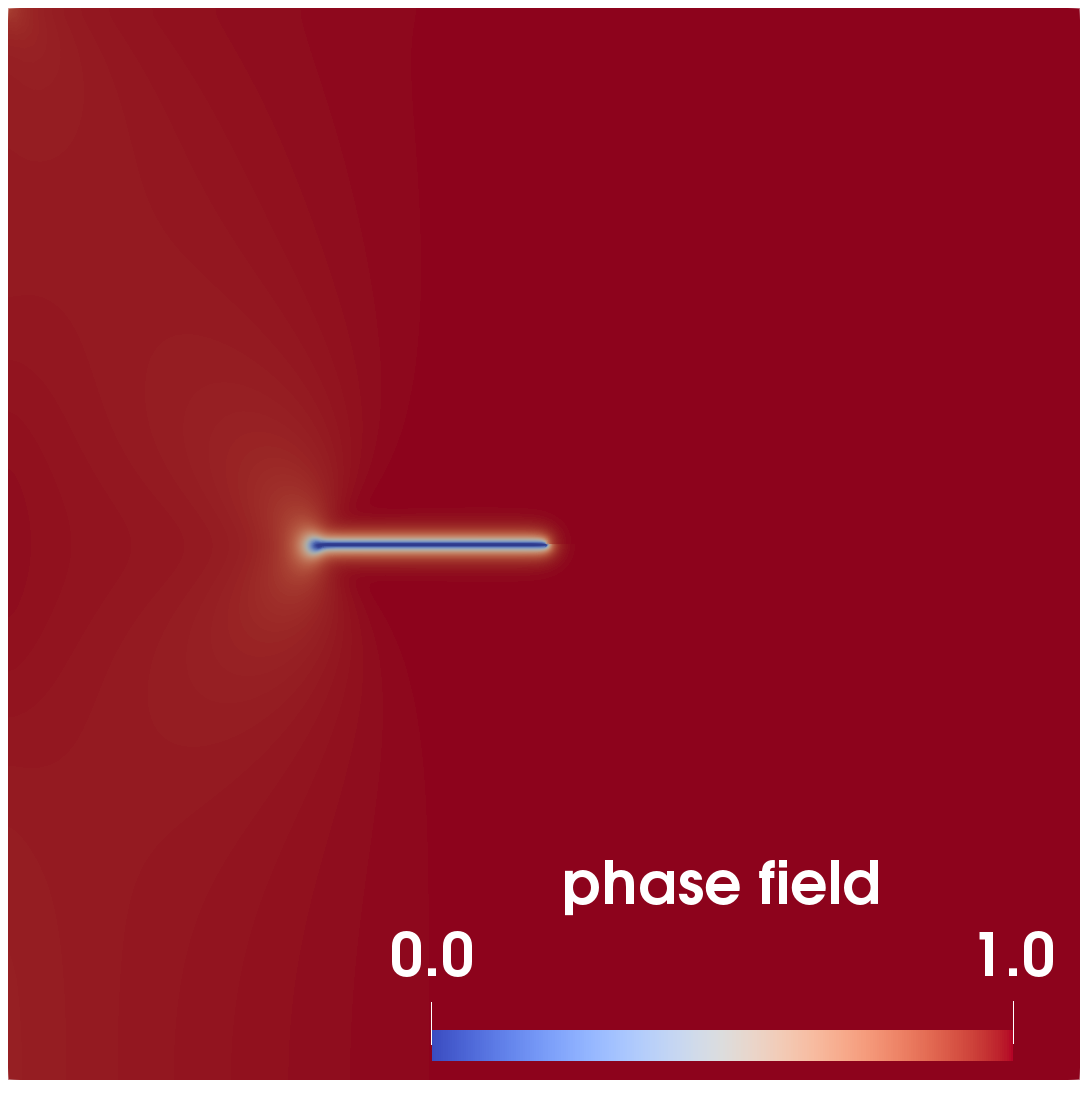}}
\hspace*{0.2in}
\subfloat[$n=30$]{\includegraphics[width= 0.3\textwidth]{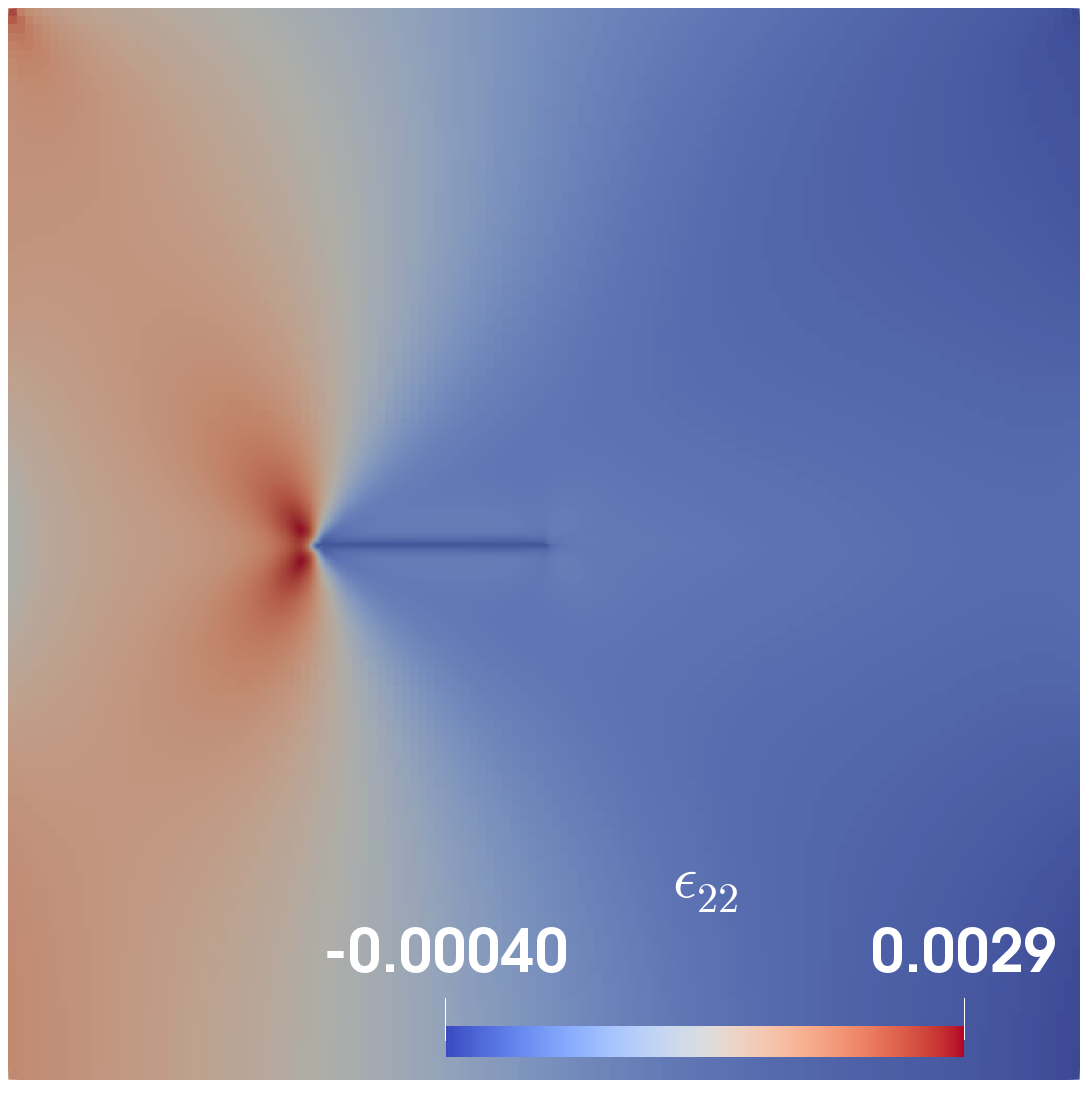}} \\
\subfloat[$n=33$]{\includegraphics[width= 0.3\textwidth]{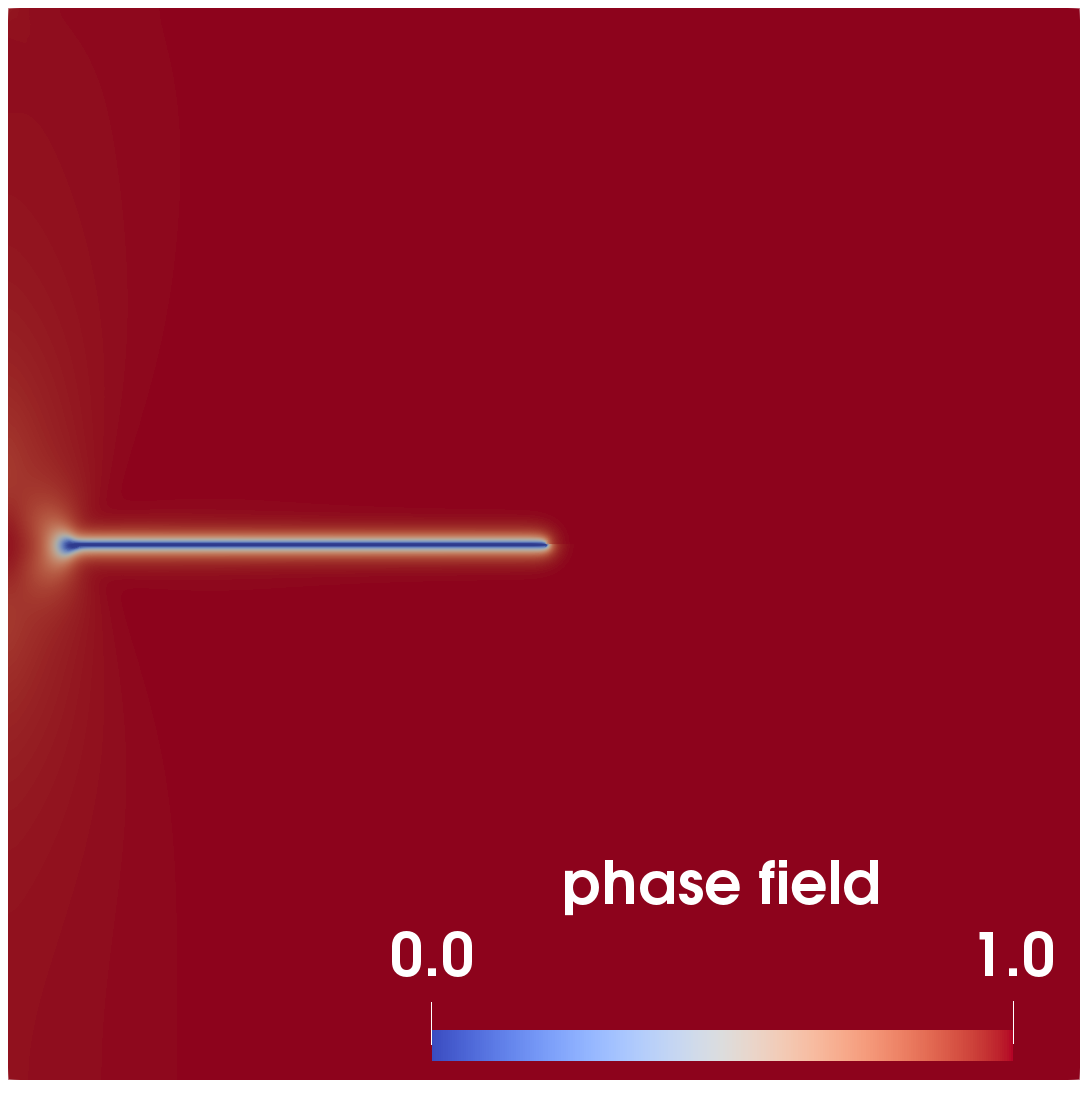}}
\hspace*{0.2in}
\subfloat[$n=33$]{\includegraphics[width= 0.3\textwidth]{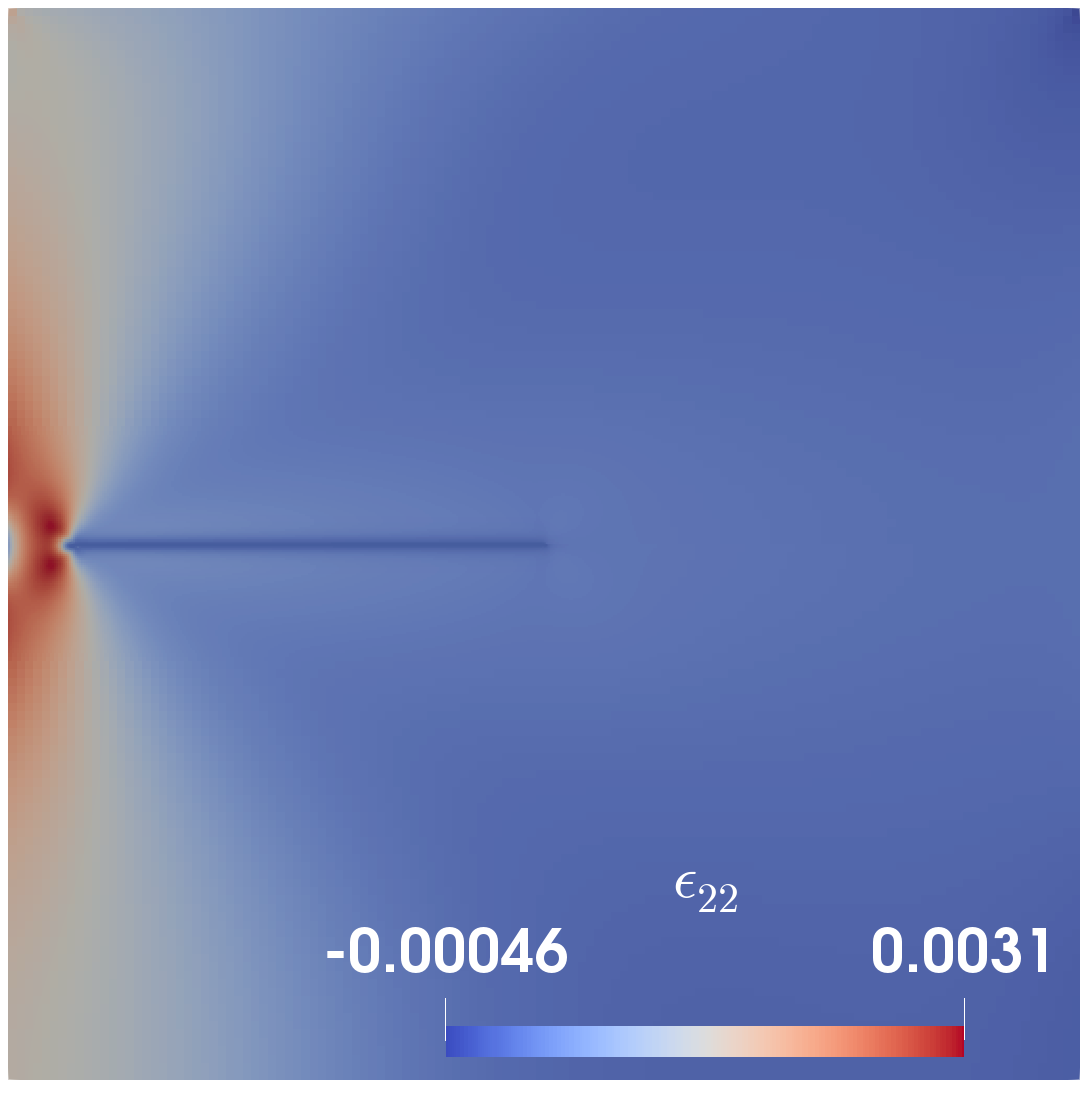}} 
\caption{ {Example 4. (Left) illustrate the phase-field values during crack evolution  for each timestep with NLSL model. 
(Right) presents the corresponding $\epsilon_{22}$ values for each case. 
The blue (lighter and thinner than LEFM) line indicates the corresponding fracture (phase-field) from the left figure. We note that 
the $\epsilon_{22}$ values are different from LEFM.}
}
\label{fig:ex4_nonlinear}
\end{figure}

First, Figure \ref{fig:ex4_linear}
and Figure \ref{fig:ex4_nonlinear} illustrate the propagation of the fracture with the phase-field values for 
LEFM and NLSL, respectively. We observe that 
NLSL model initiates the fracture earlier than LEFM.
In addition,  the overall distribution patterns of axial strain ($\varepsilon_{22}$) values are different: for LEFM, 
it is only concentrated near the vicinity of the crack-tip with 
quite larger (around 3 to 5 times) values than  NLSL.
Meanwhile, NLSL 
has more distributed values over the domain, avoiding the singularity of strain in front of the tip.

\begin{figure}[!h]
\centering
\includegraphics[width=1.0\textwidth]{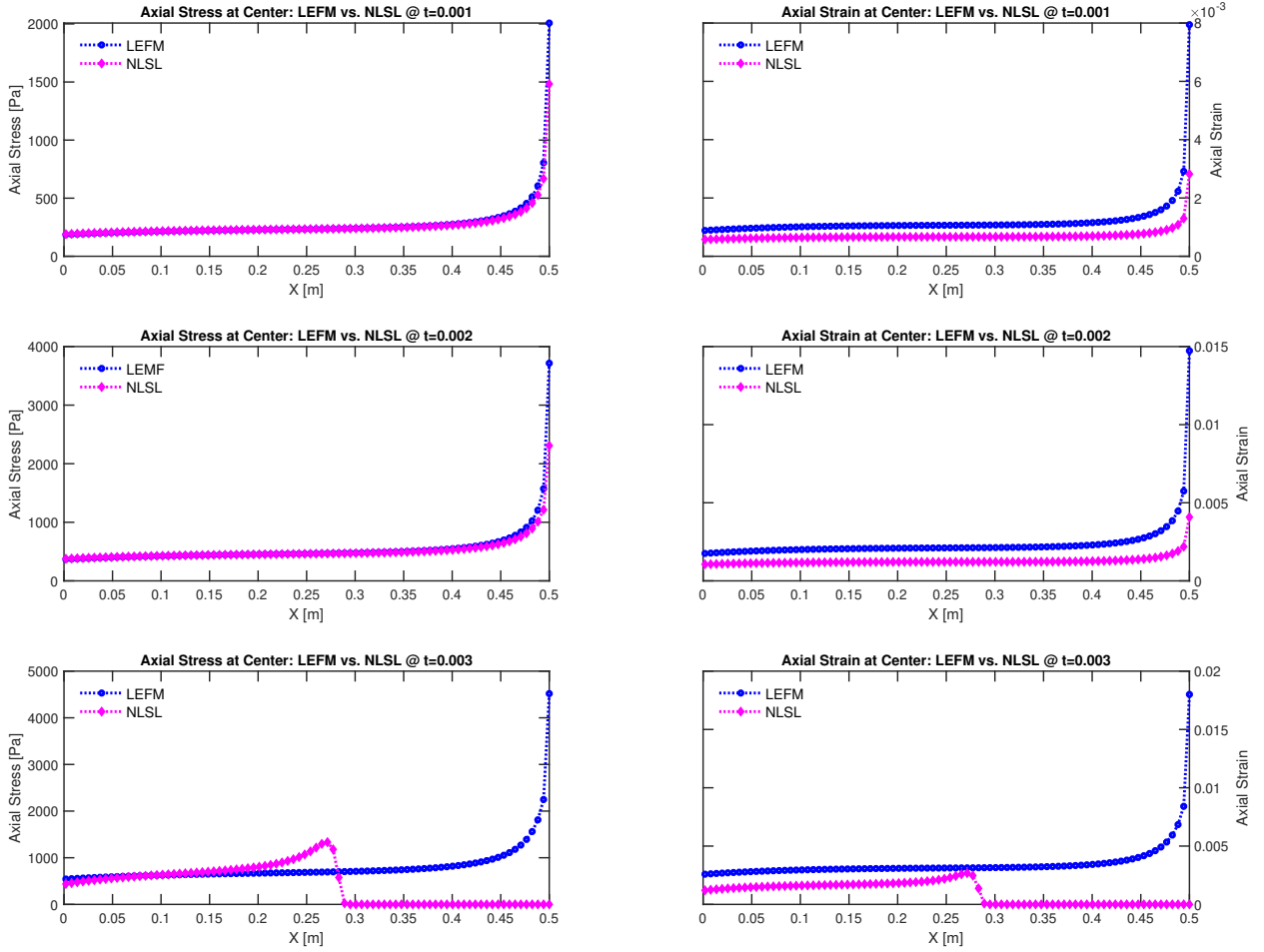}
\caption{Example 4. 
Comparisons of axial stress  $\sigma_{22}$ (Left) and strain $\epsilon_{22}$ (Right) values between {LEFM} and {NLSL} models at the time $t=0.001, 0.002$, and $0.003$.
We observe the strain-limiting effect near the tip of the fracture for 
{NLSL} model when the fracture propagation is initiated 
 {before} $t=0.003$ for {NLSL. }}
\label{fig:ex4_stress_strain}
\end{figure}

Figure~\ref{fig:ex4_stress_strain} illustrates the comparisons of the axial stress (Left) and axial strain (Right) values at the center line of $(0,0.5)-(0.5,0.5)$ between LEFM and NLSL models for three different times (snapshots) of simulations. From top row to bottom row, the timesteps of $n=10,20,$ and $30$, respectively, are presented  for axial stress ($\sigma_{22}$) and strain ($\varepsilon_{22}$) values.  
We emphasize that we observe the expected strain-limiting effects from 
NLSL model and these results also illustrate  that the 
proposed  strain-limiting model initiates the fracture propagation earlier than the linear model. 
For  NLSL model,  the crack-tip has moved forward around $n=30$ and the stress and strain values near the tip are decreased due to the crack initiation with the phase-field function.

In this example, we are also interested in the bulk (or strain) energy, the crack (or surface) energy, and  the total energy. The total energy is defined as
\begin{equation}
E_{\bfeps} := \text{Total Energy} = {\text{Bulk Energy} + \text{Surface Energy,}}
\label{ex:energy_1}
\end{equation}
and we have two different bulk energy formulations.  
For 
{LEFM}, we have 
\begin{equation}
\text{Linear Bulk Energy} := \int_\Lambda \dfrac{((1-\kappa)\varphi^2+\kappa)}{2} \left[2 \mu \bfeps(\bu) \colon \bfeps(\bu) + \lambda \;  (\nabla \cdot \bu)^2 \right] \ dx,
\label{ex:energy_2}
\end{equation}
and for 
NLSL (based on Equation~\eqref{eqn:linear_pf_new}) we have,
\textcolor{black}{\begin{equation}
\text{Nonlinear Bulk Energy} := \int_\Lambda \dfrac{((1-\kappa)\varphi^2+\kappa)}{2} \frac{\left[2 \mu \bfeps(\bu) {\colon \bfeps(\bu)} + \lambda \; { (\nabla \cdot \bu)^2}   \right]}{(1-\beta^{\alpha}|\mathbb{E}^{1/2}(\bu)|^{\alpha})^{1/\alpha}} \ dx,
\label{ex:energy_3}
\end{equation}}
where $\kappa$ is a regularization parameter taken as $\kappa=\num{e-10}h_{\min}$. 
For this example, we set {Lam\'{e}} coefficients as $\lambda=121.15~kPa,~\mu=80.77~kPa$.
Next, the crack energy is defined as
$$
 \text{Crack Energy}  := \frac{G_c}{2} \int_\Lambda \left[  \dfrac{(1-\varphi)^2}{\xi} + \xi | \nabla \varphi |^2  \right] \ dx,
$$
where $\xi = 2h_{\min}$, 
and the critical energy release rate (Griffith's criteria) is set to be $G_c=\SI{1}{\newton\per\metre}$.

\begin{figure}[!h]
\centering
\includegraphics[width=0.45\textwidth]{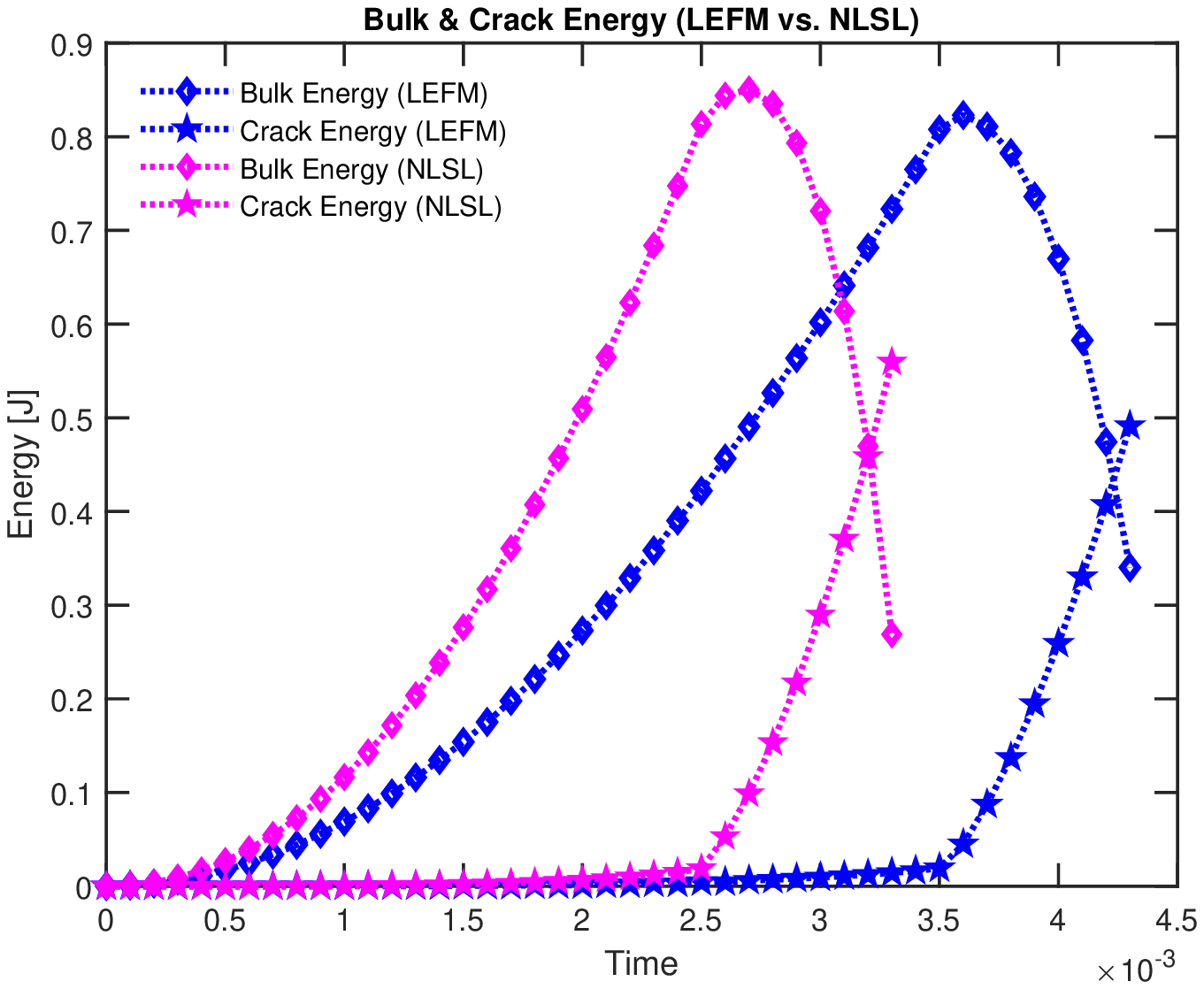}
\includegraphics[width=0.45\textwidth]{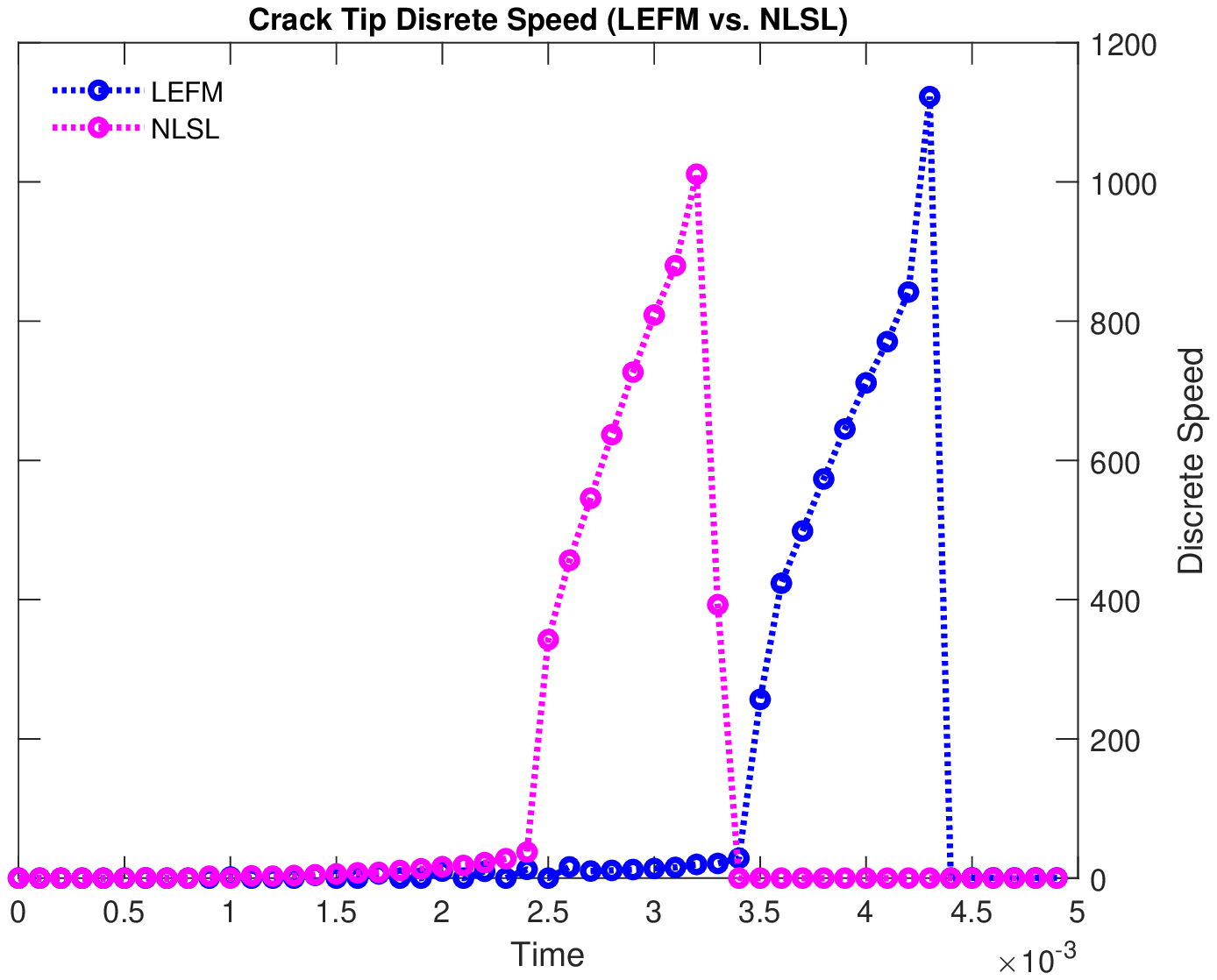}
\caption{{Example 4. (Left) Comparisons of  the bulk and crack energy between the linear and the nonlinear models. (Right) The crack (discrete) propagation rate is calculated from the crack energy.}} 
\label{fig:ex4_energy}
\end{figure}

Figure~\ref{fig:ex4_energy} (Left) presents the comparisons of bulk and crack energies following the above definitions between {LEFM} and {NLSL} models. 
Furthermore, we computed the crack growth speed as shown in 
Figure~\ref{fig:ex4_energy} (Right). The crack speed is computed using the discrete derivative of the crack (surface) energy above.
{Due to the difference of computed energies, we observe that the nonlinear strain-limiting model provides the earlier initiation of the fracture, earlier take-off, and overall larger acceleration than {LEFM}.}

\section{Conclusion}
In this paper, we investigated the strain-limiting nonlinear elasticity model coupled with the phase-field for the quasi-static fracture propagation.  
Newton iteration is employed for each nonlinear mechanics and phase-field equations, and a staggered iterative scheme, called the L-scheme, is utilized for the coupling of the system. 
Augmented Lagrangian method is employed for the constrained minimization problem with the irreversibility condition. 
Several numerical results including propagating fractures illustrate the performance of our algorithm with the capabilities of the computational framework. 
It is shown that using the proposed strain-limiting model to 
model any bulk material guarantees to bound the strain values even with the singular stress values near the crack-tip. 
Although the presented strain-limiting model requires a careful selection for the parameters $\alpha$ and $\beta$, any reasonable choice can illustrate the desired limited strain. 
Extending the current strain-limiting model to consider more freedom for the choice of the parameters is an ongoing work.

\section*{Acknowledgements}

This research done by S.~Lee is based upon work supported by the National Science Foundation under Grant No. (NSF  DMS-1913016). Other authors, Hyun C. Yoon and S. M. Mallikarjunaiah,  would like to thank the support of College of Science \& Engineering, Texas A\&M University-Corpus Christi.

\bibliographystyle{abbrv}
\bibliography{NonLinear}

\end{document}